# A Comprehensive Analysis of Insight-HXMT Gamma-Ray Burst Data. I. Power Density Spectrum

Zi-Min Zhou[1,2], Xiang-Gao Wang[1,2], En-Wei Liang[1,2], Jia-Xin Cao[1,2], Hui-Ya Liu[1,2], Cheng-Kui Li[3], Bing Li[3], Da-Bin Lin[1,2], Tian-Ci Zheng[1,2], Rui-Jing Lu[1,2], Shao-Lin Xiong[3], Ling-Jun Wang[3], Li-Ming Song[3], and Shuang-Nan Zhang[3]

[1] Guangxi Key Laboratory for Relativistic Astrophysics, School of Physical Science and Technology, Guangxi University, Nanning 530004, People's Republic of China; wangxg@gxu.edu.cn
[2] GXU-NAOC Center for Astrophysics and Space Sciences, Nanning 530004, People's Republic of China
[3] Key Laboratory of Particle Astrophysics, Institute of High Energy Physics, Chinese Academy of Sciences, Beijing 100049, People's Republic of China; zhangsn@ihep.ac.cn

## Abstract

The power density spectrum (PDS) is a powerful tool to study light curves of gamma-ray bursts (GRBs). We show the average PDS and individual PDS analysis with GRB data from the Hard X-ray Modulation Telescope (also named Insight-HXMT). The values of the power-law index of the average PDS ($\alpha_{\bar{P}}$) for long GRBs (LGRBs) vary from 1.58 to 1.29 (for 100–245, 245–600, and 600–2000 keV). The Insight-HXMT data allow us to extend the energy of the LGRBs up to 2000 keV, and a relation between $\alpha_{\bar{P}}$ and energy $E$ is obtained: $\alpha_{\bar{P}} \propto E^{-0.09}$ (8–2000 keV). We first systematically investigate the average PDS and individual PDSs for short GRBs (SGRBs), and obtain $\alpha_{\bar{P}} \propto E^{-0.07}$ (8–1000 keV), where the values of $\alpha_{\bar{P}}$ vary from 1.86 to 1.34. The distribution of the power-law index of an individual PDS ($\alpha$) of an SGRB is consistent with that of an LGRB, and the $\alpha$ value for the group with a dominant timescale (the bent power law) is higher than that for the group with no dominant timescale (the single power law). Both LGRBs and SGRBs show similar $\alpha$ and $\alpha_{\bar{P}}$, which indicates that they may be the result of similar stochastic processes. Typical values of the dominant timescale $\tau$ for LGRBs and SGRBs are 1.58 s and 0.02 s, respectively. It seems that $\tau$ varies in proportion to the duration of GRBs $T_{90}$, with a relation $\tau \propto T_{90}^{0.86}$. The GRB light curve may result from superposing a number of pulses with different timescales. No periodic or quasi-periodic signal above the $3\sigma$ significance threshold is found in our sample.

*Unified Astronomy Thesaurus concepts:* Gamma-ray bursts (629)

## 1. Introduction

Gamma-ray bursts (GRBs) are the most powerful electromagnetic explosions in the Universe (Kumar & Zhang 2015; Zhang 2018). Based on their duration ($T_{90}$), GRBs can be generally divided into two groups: short GRBs (SGRBs, $T_{90} < 2$ s) and long GRBs (LGRBs, $T_{90} > 2$ s) (Kouveliotou et al. 1993). SGRBs are associated with mergers of compact objects such as neutrons stars and black holes (Paczynski 1986, 1991; Eichler et al. 1989; Zhang et al. 2006), and LGRBs are believed to arise from the deaths of some special massive stars (Narayan et al. 1992; Woosley 1993; macFadyen & Woosley 1999; Fruchter et al. 2006; Zhang et al. 2006). Usually, the afterglow is interpreted with the external shock model (Mészáros & Rees 1997; Sari et al. 1998; Wang et al. 2015), and the prompt gamma-ray emission is interpreted within internal dissipation processes. The origin of GRB prompt emission remains not fully understood after more than 40 years of observations. The main uncertainties lie in several open questions, e.g., what is the jet composition? Where does the energy dissipation takes place? What is the radiation mechanism? (Zhang et al. 2014). The temporal behavior of GRBs is one of the critical features for prompt emission. Within the GRB prompt emission, the variabilities of the light curve are ruled by the radiation history of the jet. In principle, characterization of variability can help to constrain the radiation mechanism and dissipation processes of GRB (Ghisellini 2011; Zhang et al. 2011; Guidorzi et al. 2012).

The variability of light curves for GRBs has been characterized by several approaches, which include structure function analysis (Hook et al. 1994; Trevese et al. 1994; Cristiani et al. 1996; Aretxaga et al. 1997) and autocorrelation function analysis (Link et al. 1993; Fenimore et al. 1995; Borgonovo 2004). The power density spectrum (PDS) also provides important clues to GRB prompt emission. Beloborodov et al. (1998) applied Fourier analysis to 214 LGRBs. Their average PDS follows a power law with an index 5/3 from a few $10^{-2}$ to ~1 Hz, which coincides with the Kolmogorov spectrum of velocity fluctuations in a turbulent medium. The average PDS of GRBs is easier to characterize in terms of a general stochastic process. It provides evidence that the diversity of LGRBs is due to random realizations of the same process. The average PDSs detected by different satellites have been studied for more than a decade (Ryde et al. 2003; Guidorzi et al. 2012, 2016; Dichiara et al. 2013, 2016). Moreover, the power-law index varies from ~−1.5 to ~2 depending on the photon energy, with steeper slopes corresponding to softer energy bands. The slope of the average PDS $\alpha_{\bar{P}}$ would be smaller as the energy band increases, i.e., $\alpha_{\bar{P}} \propto E^{-0.09}$ (Dichiara et al. 2013). The individual PDS provides links between PDS properties and other individual GRB observations. Guidorzi et al. (2016) characterized individual PDSs of 215 bright LGRBs detected with Swift/BAT to probe the variety of stochastic processes taking place during the prompt emission of gamma rays. The distribution of the slope of individual PDSs is also studied and varies in a large range of 1–4 (Giblin et al. 1998; Dichiara et al. 2016).

The Hard X-ray Modulation Telescope (HXMT, also named Insight-HXMT), was launched on 2017 June 15. Insight-HXMT carries three main payloads onboard: the High Energy X-ray telescope (HE), the Medium Energy X-ray telescope (ME), and the Low Energy X-ray telescope (LE). The HE consists of 18 cylindrical NaI(Tl)/CsI(Na) (hereafter NaI, CsI) phoswich detectors (Zhang et al. 2014; Li et al. 2018). Owing to the innovative operation of the anticoincidence detectors of HE, the CsI of HE can be used for monitoring gamma-ray bursts. HE has two working modes with different detection energy ranges. Normally, it operates in the normal gain mode (Normal mode), in which CsI detectors work in the energy range 80–800 keV. To improve the detection capability for hard GRBs, the high voltage of HE can be reduced so that the measured energy range of CsI goes up to 0.2–3 MeV (GRB mode). The effective area of Insight-HXMT for typical incident geometry in both Normal and GRB modes is ∼1000 cm$^2$. This means that Insight-HXMT is one of the largest GRB detectors at energies from hundreds of keV to tens of MeV.

Thanks to the large effective area of Insight-HXMT/HE, it can detect more photons at 0.2–3 MeV than other satellites (Li et al. 2018; Luo et al. 2020), which provides data for temporal and spectral analysis to potentially reveal the nature of GRB prompt emission. We perform statistical studies of GRB prompt emission with GRB data from Insight-HXMT, which will be presented in a series of papers. In this first paper, we aim to study the average and individual PDSs with the new data sets from the Insight-HXMT/HE. Insight-HXMT/HE also allows us to study GRB PDSs with a broad energy band. In Section 2, we introduce the data analysis method of Insight-HXMT and calculation of the PDS. The results are reported in Section 3. Finally, we discuss our results in Sections 4 and 5.

## 2. Data Analysis

### 2.1. Sample Selection

We systematically investigate the Insight-HXMT GRBs from the launch of Insight-HXMT to 2021 February 25 in the Insight-HXMT catalog (Song et al. 2022). A sample of ∼248 GRBs is obtained from the Insight-HXMT data archive.[4] We extract the light curves of Insight-HXMT GRBs from the high-energy events file (hereafter HE-EVT file) included in Insight-HXMT data files.[5] The HE-EVT file is used to record information on incident photon events, including detector working mode (Normal mode or GRB mode), detector IDs (Det_ID, e.g., 0–17), channels (PI), GRB trigger time ($t_0$), photon arrival time (Time, which is recorded using *tt time system* that began on 2012 January 1), and the pulse widths (Pulse_Width). To extract the light curves, after given a start time $t_1$ and an end time $t_2$, we reject photons outside the range from $t_1$ to $t_2$. The selection of $t_1$ and $t_2$ is based on triple $T_{5\sigma}$ (Guidorzi et al. 2012, 2016; Dichiara et al. 2016). The duration $T_{5\sigma} = t_2^{5\sigma} - t_1^{5\sigma}$, where $t_1^{5\sigma}$ and $t_2^{5\sigma}$ correspond to the earliest and latest times whose counts exceed the $5\sigma$ threshold above background, respectively. To select the CsI photons for GRBs

---

[4] Insight-HXMT GRB data can be downloaded from http://archive.hxmt.cn/grb.
[5] Structure of the data file can be known from http://hxmtweb.ihep.ac.cn/u/cms/www/202011/GRBanatools_manual_beta1.4.pdf.

**Table 1**
Insight-HXMT Sample Used in PDS Analysis

| Name | $t_1$[a] (s) | $t_2$[a] (s) | $C_p$[b] | log $B_p$[c] | $T_{90}$[d] (s) |
|---|---|---|---|---|---|
| HEB170626400 | −3.264 | 8.640 | 180.82 | 0.45 | 12.10 |
| HEB170726793 | −12.928 | 24.512 | 194.75 | 0.96 | 23.70 |
| HEB170817908 | −2.304 | 5.568 | 266.20 | 0.14 | 2.64 |
| HEB170826818 | −8.384 | 17.728 | 221.79 | 0.75 | 10.12 |
| HEB170923188 | −12.032 | 22.720 | 166.45 | 1.13 | 12.41 |
| HEB170926340 | −1.232 | 2.768 | 194.92 | 0.14 | 2.88 |
| HEB171210492 | −9.920 | 22.336 | 106.57 | 1.62 | 20.20 |
| HEB171215705 | −16.000 | 17.600 | 136.51 | 1.33 | 32.20 |
| HEB180103047 | −13.952 | 31.552 | 222.05 | 0.58 | 147.80 |
| HEB180202211 | −6.784 | 13.952 | 216.85 | 0.63 | 11.70 |
| HEB180210517 | −24.472 | 48.890 | 209.07 | 1.75 | 26.70 |
| HEB180221520 | −20.864 | 46.144 | 159.04 | 1.53 | 65.80 |
| HEB180305393 | −10.688 | 19.456 | 699.61 | 0.01 | 7.20 |
| HEB180326143 | −10.112 | 28.480 | 356.62 | 0.60 | 14.10 |
| HEB180404091 | −21.248 | 37.312 | 111.84 | 1.88 | 16.60 |
| HEB180411359 | −90.111 | 185.600 | 102.10 | 2.39 | 92.10 |
| HEB180718762 | −18.752 | 38.848 | 207.32 | 1.26 | 27.30 |
| HEB180804554 | −29.312 | 34.240 | 138.01 | 1.65 | 86.94 |
| HEB180804930 | −4.160 | 8.320 | 170.58 | 0.81 | 4.99 |
| HEB180828789 | −6.336 | 16.704 | 386.39 | 0.18 | 7.61 |
| HEB180925609 | −9.536 | 19.264 | 172.15 | 1.06 | 23.23 |
| HEB181014479 | −30.234 | 30.620 | 136.78 | 1.78 | 17.64 |
| HEB181201111 | −60.730 | 76.180 | 546.21 | 0.80 | 15.18 |
| HEB181212692 | −3.520 | 8.192 | 380.72 | 0.09 | 5.48 |
| HEB190103877 | −10.112 | 22.720 | 112.80 | 1.53 | 18.73 |
| HEB190110725 | −1.776 | 2.224 | 117.17 | 0.54 | 7.19 |
| HEB190117608 | −18.624 | 37.632 | 420.18 | 0.57 | 23.29 |
| HEB190131964 | −19.520 | 39.616 | 110.27 | 1.72 | 23.40 |
| HEB190215771 | −10.112 | 20.992 | 148.21 | 1.16 | 14.79 |
| HEB190321931 | −1.039 | 2.961 | 209.55 | −0.04 | 5.08 |
| HEB190324348 | −40.256 | 80.704 | 152.11 | 1.82 | 46.07 |
| HEB190326313 | −52.224 | 103.104 | 186.14 | 1.78 | 58.45 |
| HEB190531840 | −39.616 | 77.696 | 674.94 | 0.52 | 27.82 |
| HEB190604446 | −3.712 | 11.072 | 127.59 | 1.06 | 6.37 |
| HEB190605110 | −9.216 | 18.816 | 142.93 | 1.25 | 14.69 |
| HEB190613449 | −2.752 | 5.888 | 202.11 | 0.35 | 3.59 |
| HEB190615612 | −0.704 | 3.904 | 134.16 | 0.41 | 23.69 |
| HEB190620507 | −27.136 | 41.408 | 247.13 | 1.09 | 50.26 |
| HEB190806675 | −11.200 | 22.208 | 100.28 | 1.61 | 31.55 |
| HEB190828783 | −3.904 | 7.808 | 197.74 | 0.72 | 5.21 |
| HEB190928551 | −34.176 | 40.512 | 1212.79 | −0.23 | 13.20 |
| HEB191019970 | −0.768 | 27.648 | 106.60 | 1.70 | 34.50 |
| HEB191019971 | −26.240 | 39.232 | 131.97 | 1.66 | 26.19 |
| HEB191202867 | −1.344 | 2.688 | 101.33 | 0.64 | 11.09 |
| HEB191221860 | −13.824 | 22.272 | 338.42 | 0.53 | 8.85 |
| HEB191227069 | −26.816 | 39.808 | 992.98 | −0.13 | 20.65 |
| HEB200111632 | 3.856 | 7.856 | 166.63 | 0.39 | 8.47 |
| HEB200125863 | −3.008 | 7.168 | 564.05 | −0.37 | 3.84 |
| HEB200219998 | −3.264 | 5.184 | 124.38 | 0.98 | 5.88 |
| HEB200227305 | −6.272 | 9.664 | 101.53 | 1.33 | 19.45 |
| HEB200412381 | −8.832 | 14.784 | 620.29 | −0.02 | 5.51 |
| HEB200413712 | −1.712 | 2.288 | 123.71 | 0.56 | 3.98 |
| HEB200619108 | −10.368 | 20.736 | 179.57 | 1.25 | 18.49 |
| HEB200716956 | −1.933 | 4.487 | 692.09 | −0.68 | 2.16 |
| HEB201013157 | −7.355 | 14.244 | 698.34 | −0.21 | 6.77 |
| HEB210112068 | −10.662 | 20.857 | 160.83 | 1.17 | 12.27 |
| HEB210213286 | −61.320 | 76.650 | 150.16 | 1.50 | 15.33 |
| HEB190906767 | −4.462 | 8.786 | 254.77 | 0.33 | 16.71 |
| HEB191218112 | −48.203 | 94.837 | 204.95 | 1.62 | 49.81 |
| HEB201105229 | −0.978 | 6.702 | 180.51 | 0.46 | 8.46 |
| HEB170906029 | 24.128 | 41.792 | 64.77 | 1.12 | 64.60 |
| HEB180111695 | −12.735 | 25.472 | 76.61 | 1.29 | 19.94 |
| HEB210121779 | −39.675 | 79.363 | 334.82 | 0.93 | 13.24 |
| HEB210207911 | 0.084 | 4.084 | 64.40 | 0.87 | 2.03 |

**Table 1**
(Continued)

| Name | $t_1$[a] (s) | $t_2$[a] (s) | $C_p$[b] | $\log B_p$[c] | $T_{90}$[d] (s) |
|---|---|---|---|---|---|
| HEB190906045 | −0.071 | 0.142 | 51.07 | −0.07 | 0.07 |
| HEB190903721 | −0.514 | 0.628 | 21.16 | 1.54 | 0.11 |
| HEB190724030 | −0.060 | 0.120 | 32.87 | 0.44 | 0.06 |
| HEB190610477 | 0.431 | 2.138 | 44.08 | 0.94 | 0.57 |
| HEB190326316 | −0.018 | 0.036 | 42.96 | −0.36 | 0.02 |
| HEB180618030 | −0.420 | 0.880 | 22.31 | 1.63 | 0.24 |
| HEB180402406 | −0.171 | 0.342 | 27.32 | 1.01 | 0.17 |
| HEB171223818 | −0.170 | 0.340 | 61.04 | 0.22 | 0.17 |
| HEB171030728 | 0.040 | 0.190 | 18.20 | 0.78 | 0.05 |
| HEB170802637 | −0.977 | 1.954 | 41.49 | 1.37 | 0.98 |
| HEB170801208 | −0.029 | 0.078 | 24.99 | 0.20 | 0.03 |

**Notes.**
[a] The PDS is calculated in the time interval from $t_1$ to $t_2$, where $t_1 = t_1^{5\sigma} - T_{5\sigma}$ and $t_2 = t_2^{5\sigma} + T_{5\sigma}$.
[b] Net peak counts of the GRB.
[c] The predicted value of Poisson level $B_p = 2V/C_p^2$ from Equation (2).
[d] $T_{90}$ is derived from the first and last bins with 5% and 95% total net counts of a GRB (Koshut et al. 1996).

(Xiao et al. 2020), pulse shape discrimination technology was employed to distinguish which phoswich detected the arrival photon (Liu et al. 2020). Here, we can select the CsI photons with the criterion that Pulse_Width should be larger than 70 (Liu et al. 2020). The deposited energy covered by Insight-HXMT is divided into 256 channels. The deposited energy ($E$) can be expressed by a linear energy–channel relation, $E = a \times \mathrm{PI} + b$, where $a$ and $b$ are constants given in Luo et al. (2020). To obtain the net counts $C(t)$ for the PDS calculation, we fitted background counts $C_B(t)$ using a first- or second-order polynomial. Then $C(t)$ was obtained by subtracting $C_B(t)$ from the total counts of the original light curve $C_O(t)$ (e.g., Beloborodov et al. 2000; Dichiara et al. 2013).

The energy bands selected are 100–600 keV and 400–2000 keV for Normal mode and GRB mode, respectively. For LGRBs, we use the light-curve data in a time bin of 64 ms, which is convenient to compare with the results of other satellites selected in 64 ms bins in previous works (Guidorzi et al. 2012, 2016; Dichiara et al. 2013, 2016). For SGRBs, we use the light-curve data also in a time bin of 2 ms. To avoid the influence of noise from the background as much as possible, we exclude weak GRBs whose net peak counts (time bin = 64 ms) are less than 100 (Beloborodov et al. 2000). In Insight-HXMT, GRBs whose count rate in the 18 CsI detectors exceeds 20,000 counts s$^{-1}$ are saturated, which leads to a loss of counts in the light curve (Song et al. 2022). The saturated GRBs (in the "IRON" sample of Song et al. 2022) are excluded. Since HEB200415366 (GRB 200415A) is identified as an extragalactic magnetar giant flare from NGC 253 (Lin et al. 2020; Yang et al. 2020; Roberts et al. 2021; Svinkin et al. 2021), it is excluded from our sample. Finally, 64 LGRBs (59 Normal mode and 5 GRB mode) and 11 SGRBs (all in Normal mode) are selected in our sample, whose information is presented in Table 1.

### 2.2. Calculation of Power Density Spectrum

The average PDS over a large number of GRBs provides a way to characterize the phenomenon in terms of a stochastic process (Beloborodov et al. 1998, 2000; Guidorzi et al. 2012; Dichiara et al. 2013). The individual PDS provides links between PDS properties and other individual GRB observations (Dichiara et al. 2013, 2016; Guidorzi et al. 2016). The power density $P$ for an individual GRB is given by (Beloborodov et al. 1998)

$$P = \frac{A_{-f}A^*_{-f} + A_f A^*_f}{2}, \quad (1)$$

where $A_f$ ($A_{-f}$) is the Fourier coefficient of a given light curve with net counts $C(t)$, and $A^*_f$ ($A^*_{-f}$) is the conjugate of $A_f$ ($A_{-f}$). The fast Fourier transform (FFT) technique is employed to calculate $A_f$ for the light curve (Jenkins & Watts 1968; Leahy et al. 1983).

The average power density $\bar{P}$ is calculated by summing up all individual PDSs and dividing by the number of GRBs in the sample (Beloborodov et al. 1998, 2000). For individual GRB, the value of $P$ is affected by the brightness. If no normalization is employed, the shapes and fluctuations of the brightest GRBs would dominate in $\bar{P}$ and the contribution of weak (less bright) GRBs may be obscured (Beloborodov et al. 1998). Therefore, it is necessary to perform a normalization in the calculation of $\bar{P}$. The PDS is usually normalized in two ways, e.g., Leahy normalization (Leahy et al. 1983) and peak normalization (Beloborodov et al. 1998). For Leahy normalization, the PDS is divided by the total variance $V$ of the corresponding light curve, and the expected value of the Poisson level is equal to 2 (Leahy et al. 1983; Guidorzi et al. 2012; Dichiara et al. 2013). For peak normalization, the PDS is divided by the net peak counts $C_p^{26}$ of the corresponding light curve, and the expected value of the Poisson level $B_p$ is (see Appendix for detailed derivations)

$$B_p = \frac{E\{P\}}{C_p^2} = \frac{2V}{C_p^2}. \quad (2)$$

The values of $C_p$ and $\log B_p$ for each burst are listed in Table 1. The average PDS results with peak normalization have been reported from different satellites, e.g., CGRO/BATSE (Beloborodov et al. 1998, 2000) and Swift/BAT (Guidorzi et al. 2012). To compare with previous results conveniently, we adopted the peak normalization in our work.

### 2.3. Fitting of Power Density Spectrum

Most of the GRB PDSs present two breaks. The first break depends on the dominant timescale of the pulses $\tau$ ($\tau = 1/2\pi f_b$, where $f_b$ is the first break frequency) (e.g., Frontera & Fuligni 1979; Belli 1992; Dichiara et al. 2013; Guidorzi et al. 2016). The second break represents the appearance of Poisson noise. Generally, the average PDS can be described as a smoothly broken power-law function plus a Poisson constant (SBPL model; e.g., Lazzati 2002; Guidorzi et al. 2012; Dichiara et al. 2013):

$$P = N\left[\left(\frac{f}{f_b}\right)^{\omega\alpha_1} + \left(\frac{f}{f_b}\right)^{\omega\alpha_2}\right]^{-1/\omega} + B, \quad (3)$$

where $N$ is the normalization constant, $f_b$ is the break frequency of the average PDS, and $B$ is the Poisson level. For frequency $f$

---
[6] The amplitude of PDS is proportional to the net peak counts $C_p$ (Beloborodov et al. 1998, 2000), i.e., $P \propto A_f^2 \propto C_p^2$.

less than and greater than $f_b$, the corresponding power density $P$ decreases with power-law indices $\alpha_1$ and $\alpha_2$ (where $\alpha_2 > \alpha_1$), respectively. $\omega$ is the smooth parameter, which is set to a fixed value, i.e., $\omega = 10$, to compare with the results of other satellites (Guidorzi et al. 2012; Dichiara et al. 2013) whose smooth parameters are also set as 10.

The PDS of a pulse with a fast rise and exponential decay is expected to be flat ($\alpha_1$ tends to 0 at $f \ll f_b$) (e.g., Lazzati 2002; Dichiara et al. 2016). For an individual PDS of a GRB, the SBPL can be simplified as a BPL (e.g., Guidorzi et al. 2016; Dichiara et al. 2016):

$$P = N\left[1 + \left(\frac{f}{f_b}\right)^\alpha\right]^{-1} + B, \qquad (4)$$

where $\alpha$ is the power-law index in the individual PDS.

For many PDSs, there is no obvious break in the red noise at low frequencies, so the $f_b$ value cannot be constrained using the BPL model. In this scenario, it can be fitted with a simple power-law function plus a Poisson constant (PL model):

$$P = Nf^{-\alpha} + B. \qquad (5)$$

The PL model may be due to the limit $f \gg f_b$ (Dichiara et al. 2016). In this work, we fit the average PDS with the SBPL model, while both BPL and PL models are applied for individual PDSs.

In addition, the Bayesian procedure is adopted (Guidorzi et al. 2016) to estimate the best-fitting parameters of models through a Markov Chain Monte Carlo (MCMC) algorithm. Bayes' theorem allows one to calculate the posterior distributions ($p(\theta|D, M)$) for model $M$ in the given data $D$,

$$p(\theta|D, M) = \frac{L(D|\theta, M)\pi(\theta|M)}{Z(D|M)}, \qquad (6)$$

where $\theta$ is the fitting parameter of model $M$, $L(D|\theta, M)$ is the likelihood (Trotta 2008; Guidorzi et al. 2016), $\pi(\theta|M)$ is the prior knowledge of the model parameter, and $Z(D|M)$ is the Bayesian evidence that represents the average likelihood for model $M$ (Trotta 2008; Sarin et al. 2019). The Python package emcee (Foreman-Mackey et al. 2013) is used for our MCMC simulation. $5 \times 10^4$ sets of parameters are generated and the initial 10% of iterations are abandoned. Based on a large number of data sets, we assume that $p(\theta|D, M)$ follows a multidimensional Gaussian distribution referred to as the Laplace approximation,[7] and its expected value can be seen as the best-fitting parameter of model $M$ (Vaughan 2010).

The selection analysis of the Bayesian model (e.g., Jeffreys 1935; Kass & Raftery 1995; Trotta 2008), is adopted to compare the BPL and PL models for an individual PDS. The ratio between the posterior probabilities of BPL and PL ($p(M_{BPL}|D)$ and $p(M_{PL}|D)$) is

$$\frac{p(M_{BPL}|D)}{p(M_{PL}|D)} = \frac{Z(D|M_{BPL})p(M_{BPL})}{Z(D|M_{PL})p(M_{PL})}, \qquad (7)$$

where $p(M)$ is the prior probability for model $M$ ($M_{BPL}$ and $M_{PL}$ correspond to the BPL and PL models, respectively). Assuming the prior probabilities for these two models are equal ($p(M_{PL}|D) = p(M_{BPL}|D)$) for the given data $D$,

$$\frac{p(M_{BPL}|D)}{p(M_{PL}|D)} = \frac{Z(D|M_{BPL})}{Z(D|M_{PL})}. \qquad (8)$$

The Bayes factor, which could be easily obtained from MCMC simulation, is defined to represent the ratio (Equation (8)) (e.g., Kass & Raftery 1995; Trotta 2008):

$$\begin{aligned} BF &\equiv \frac{Z(D|M_{BPL})}{Z(D|M_{PL})} \\ &= \frac{\int L(D|\theta_{BPL}, M_{BPL})\pi(\theta_{BPL}, M_{BPL})\, d\theta_{BPL}}{\int L(D|\theta_{PL}, M_{PL})\pi(\theta_{PL}, M_{PL})\, d\theta_{PL}}, \end{aligned} \qquad (9)$$

and

$$\log BF = \log Z(D|M_{BPL}) - \log Z(D|M_{PL}), \qquad (10)$$

where $\theta_{BPL}$ and $\theta_{PL}$ are the parameters of the BPL model and PL model, respectively. The logarithmic Bayesian factor $\log BF$ can be used for model comparison: $\log BF > 0$ indicates that the posterior probability of the BPL model is larger than that of the PL model. Hence we select the BPL model in this case. In contrast, for the case of $\log BF < 0$, we select the PL model[8] (Jeffreys 1935; Trotta 2008).

In the high-frequency region, the PDS with peak normalization tends to the Poisson level $B_p$ (Equation (2)). To check whether the fitting results of PDSs are reliable, we try to compare $B_p$ and the fitting values $B_f$. The value of the Nyquist frequency of some GRBs may be not large enough (which depends on the time bin), which leads to the Poisson noise level being buried below the red noise of the signal across the frequency range of their PDSs. Hence we make the prior distribution $\pi(B|M)$ for each GRB a Gaussian distribution, centered on its expected $B_p$ with a standard deviation of 0.1 $B_p$ (Guidorzi et al. 2016). As shown in Figure 1, the fitting results of Poisson level $B_f$ and $B_p$ are consistent with each other.

## 3. Results

The Insight-HXMT GRB detector has two main advantages: large detector area and high time resolution. The light curves of Insight-HXMT data allow us to explore higher energy and frequency in the PDS.

### 3.1. Average Power Density Spectra of LGRBs

For the LGRBs, we calculate the average PDSs in three energy bands: 100–245 keV (low energy), 245–600 keV (middle energy), and 600–2000 keV (high energy). To compare with the previous results (e.g., Beloborodov et al. 1998; Guidorzi et al. 2012; Dichiara et al. 2013), we first set the data with 64 ms binning time. The best-fitting parameters ($\log N$, $\log B$, $\alpha_1$, $\alpha_2$, and $\log f_b$) of the average PDSs for LGRBs are shown in Figure 2 and also listed in Table 2.

Among these parameters, our focus is on the power-law index $\alpha_2$, which ranges from a few 0.01 Hz to several Hz,

---

[7] Noting that the assumption is rough for some cases, the posteriors may not be well represented by a multidimensional Gaussian distribution, although many promising improvements have been suggested (e.g., DiCiccio et al. 1997; Weinberg 2012).

[8] Previous results have shown that there are some individual PDSs of GRBs that do not show clear breaks in the low-frequency region, which may be due to the superposition of PDSs of multiple pulse components (the group with no dominant timescale, described in Section 4 or in Dichiara et al. 2016). Therefore, instead of "null hypothesis model," we select the fitting model simply based on the larger value of posterior probability.

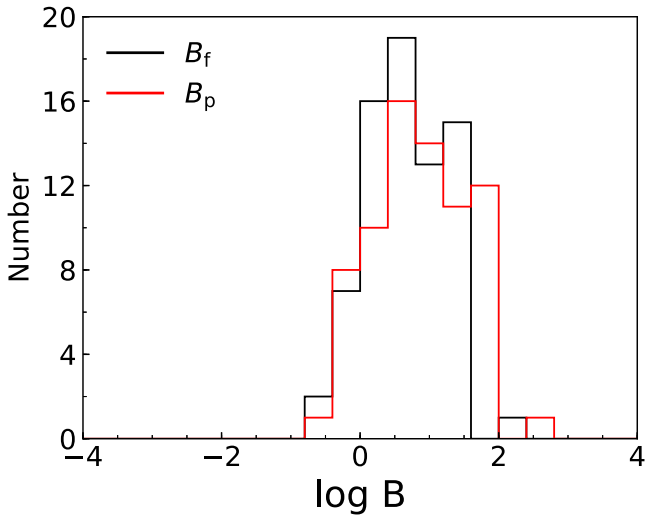

**Figure 1.** The distribution of the logarithmic Poisson level with the predicted values $B_p$ (red) and fitting values $B_f$ (black).

which was also the main focus in previous investigations (Beloborodov et al. 1998; Guidorzi et al. 2012; Dichiara et al. 2013). We denote $\alpha_2$ as $\alpha_{\bar{P}}$ for the average PDS hereafter. The value of $\alpha_{\bar{P}}$ is shallower as energy increases, being $1.58^{+0.06}_{-0.06}$, $1.37^{+0.07}_{-0.06}$, and $1.29^{+0.17}_{-0.17}$ for the three energy channels (low, middle, and high energy), respectively. This indices of the spectra are similar to the previous values in different satellites, e.g., Swift, BATSE, and Fermi (Guidorzi et al. 2012; Dichiara et al. 2013). A second break of the average PDS at 1–10 Hz has been found in CGRO/BATSE, Fermi/GBM, and BeppoSAX/GRBM data (Beloborodov et al. 2000; Dichiara et al. 2013). We extract the Insight-HXMT/HE light curves with 8 ms and 0.2 ms time resolutions, which allows us to explore the behavior of the average PDS up to 2500 Hz (Nyquist frequency, $1/(2 \times 0.0002 \text{ s})$). We rebin the average PDS and subtract the Poisson noise $B$ to find whether a break exists at a higher frequency. For LGRBs, we rebin the PDSs with a step $\Delta \log(f/\text{Hz}) = 0.25$ at $f > 4 \times 10^{-2}$ Hz. At $f \leqslant 4 \times 10^{-2}$ Hz, we set the uniform step $\Delta \log(f/\text{Hz}) = 0.4$, due to insufficient data points at lower frequencies. In the high-frequency region ($f > 5$ Hz), we rebin the PDSs with a step $\Delta \log(f/\text{Hz}) = 0.9$. The average noise-subtracted PDSs with different time resolutions are shown in Figure 3. At the higher frequency of the average PDS of Insight-HXMT GRBs, no obvious break exists. It is noted that the values of average power density from 100 to 2500 Hz have large uncertainties. We determine no break exists up to a frequency of $\sim$10 Hz. The results are consistent with the results of Swift/BAT (Guidorzi et al. 2012).

### 3.2. Average Power Density Spectra of SGRBs

The average PDSs of Insight-HXMT SGRBs are calculated in two energy bands, 100–245 and 245–600 keV. To ensure the quality of light curves in our sample, we set the SGRB data with 2 ms binning time after comparing different time bins (1 ms, 2 ms, 4 ms, and 8 ms). Figure 4 and Table 2 show the average PDSs of SGRBs and their best-fitting parameters ($\log N$, $\log B$, $\alpha_1$, $\alpha_2$, and $\log f_b$). In order to explore whether the break at higher frequency exists in PDSs of SGRBs, the average noise-subtracted PDSs of SGRBs are shown in Figure 5. Since the sample size of SGRBs is less than that of LGRBs, we only calculate the average PDS of Insight-HXMT Normal mode SGRBs with a 2 ms time bin. For SGRBs, we rebin the PDSs of SGRBs with a step $\Delta \log(f/\text{Hz}) = 0.3$ at $f \leqslant 4$ Hz and $\Delta \log(f/\text{Hz}) = 0.5$ at $f > 4$ Hz. The index of the average PDS $\alpha_{\bar{P}}$ ($\alpha_2$) of SGRBs with Insight-HXMT is $1.52^{+0.55}_{-0.36}$ for 100–245 keV and $1.48^{+0.58}_{-0.44}$ for 245–600 keV. We do not observe any obvious break of average power density up to a frequency of $\sim$100 Hz. It is noted that, although the light curves of SGRBs with a 2 ms bin make the frequency of the average PDS extend to 250 Hz ($1/(2 \times 0.002 \text{ s})$), we cannot determine whether a break exists at a frequency of several hundred Hz, since the values of average power density at several hundred Hz have larger uncertainties.

### 3.3. Individual Power Density Spectra of LGRBs

The average PDS represents the general behavior from a large number of GRBs, which could not directly probe the variety of stochastic processes taking place during the gamma-ray prompt emission. Generally, a light curve is a result of superposing a number of pulses with different timescales (Dichiara et al. 2016). Based on Swift GRB data, Dichiara et al. (2016) and Guidorzi et al. (2016) classified the individual GRB PDSs into two types. The first type is dominated by specific timescales, resulting in a break with frequency $f_b$, and can be fitted with the BPL model. The other type may have several different timescales with similar weights, which result in no distinct break in the PDS and can be well fitted with the PL model.

To investigate the individual properties of GRBs, we also calculate the PDS for individual Insight-HXMT GRBs with the BPL and PL models, respectively. The best-fitting model is selected with the Bayesian model selection (as described in Section 2.2). Furthermore, the goodness of the best-fitting models is also assessed with the p-value ($p_{KS}$) associated with the Kolmogorov–Smirnov (K-S) test, to compare the fitting results with $\chi^2_2$ distribution. Figure 6 illustrates examples for the individual PDSs of LGRBs. The results are listed in Table 3 including $\log BF$, $p_{KS}$, $\log N$, $\log B$, $\alpha$, and $\log f_b$. For 39 and 25 LGRBs, the best-fitting models are BPL and PL models, respectively. The mean value of $p_{KS}$ for all LGRBs is $0.46 \pm 0.16$, and the values of $p_{KS}$ for individual GRB PDSs are all larger than 0.05.

Figure 7 and Table 4 display the distributions of $\alpha$ of LGRBs for both BPL and PL models. The typical values of $\alpha$ (mean values with their $1\sigma$ uncertainty, hereafter) for the BPL sample and PL sample are $2.49 \pm 0.78$ and $1.95 \pm 0.74$ respectively. The $\alpha$ value of the BPL is larger than that of the PL, which is similar to the results of Guidorzi et al. (2016).

### 3.4. Individual Power Density Spectra of SGRBs

We also investigate the individual PDSs for SGRBs. We use the Bayesian model selection to select the favored fitting model. The typical value of $p_{KS}$ is $0.61 \pm 0.29$ and all the values of $p_{KS}$ significantly support the null hypothesis that the individual PDS with its favored model is consistent with the $\chi^2_2$ distribution. Figure 8 illustrates examples for the individual PDSs of SGRBs and their best-fitting models. The fitting results ($\log N$, $\log B$, $\alpha$, and $\log f_b$) are listed in Table 5. The typical values of $\alpha$ of the BPL and PL samples are

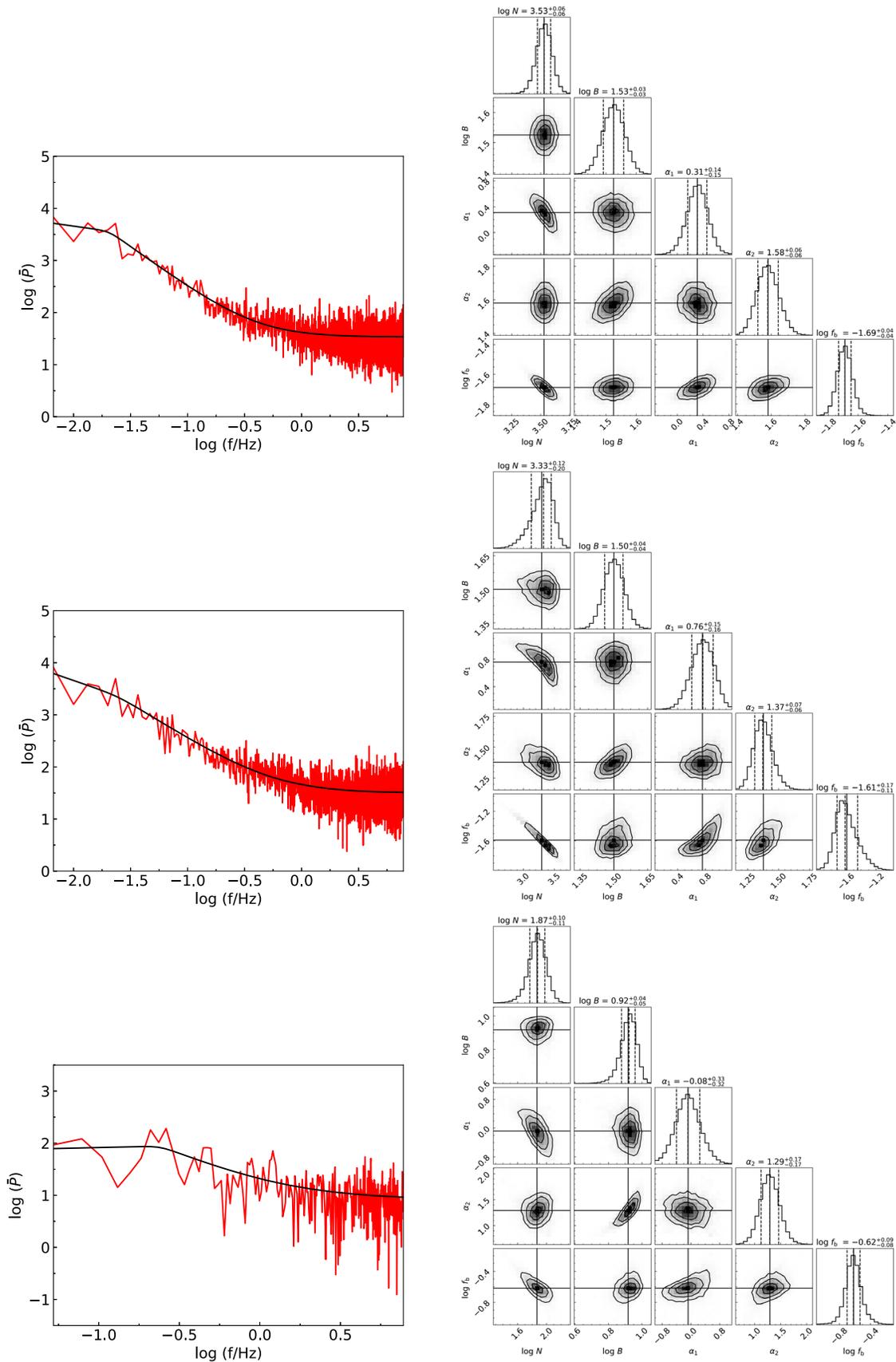

**Figure 2.** Fitting results of the average PDS for LGRBs with 64 ms binning time. The left panels are the average PDSs of LGRBs, and the black solid lines show the corresponding best-fitting model. The right panels are the distributions of parameters $\log N$, $\log B$, $\alpha_1$, $\alpha_2$, and $\log f_b$. The top, middle, and bottom panels correspond to energy bands of 100–245, 245–600, and 600–2000 keV.

Table 2
Fitting Results of the Average PDSs

| Type[a] | Energy Range (keV) | log N | log B | $\alpha_1$ | $\alpha_2$ | log $f_b$[b] (Hz) |
|---|---|---|---|---|---|---|
| LGRB | 100–245 | $3.53^{+0.06}_{-0.06}$ | $1.53^{+0.03}_{-0.03}$ | $0.31^{+0.14}_{-0.15}$ | $1.58^{+0.06}_{-0.06}$ | $-1.69^{+0.04}_{-0.04}$ |
| LGRB | 245–600 | $3.33^{+0.12}_{-0.20}$ | $1.50^{+0.04}_{-0.04}$ | $0.76^{+0.15}_{-0.16}$ | $1.37^{+0.07}_{-0.06}$ | $-1.61^{+0.17}_{-0.11}$ |
| LGRB | 600–2000 | $1.87^{+0.10}_{-0.10}$ | $0.92^{+0.04}_{-0.05}$ | $-0.08^{+0.33}_{-0.32}$ | $1.29^{+0.17}_{-0.17}$ | $-0.62^{+0.09}_{-0.08}$ |
| SGRB | 100–245 | $2.32^{+0.24}_{-0.39}$ | $1.52^{+0.05}_{-0.06}$ | $0.59^{+0.31}_{-0.78}$ | $1.52^{+0.55}_{-0.36}$ | $0.40^{+0.35}_{-0.25}$ |
| SGRB | 245–600 | $2.30^{+0.23}_{-0.37}$ | $1.76^{+0.05}_{-0.06}$ | $0.54^{+0.34}_{-0.69}$ | $1.48^{+0.58}_{-0.44}$ | $0.40^{+0.33}_{-0.27}$ |
| SGRB | 15–50 (Swift) | $2.26^{+0.14}_{-0.20}$ | $1.10^{+0.01}_{-0.01}$ | $0.87^{+0.16}_{-0.24}$ | $1.86^{+0.16}_{-0.13}$ | $0.13^{+0.13}_{-0.09}$ |
| SGRB | 50–150 (Swift) | $2.32^{+0.17}_{-0.20}$ | $1.10^{+0.01}_{-0.01}$ | $0.88^{+0.15}_{-0.22}$ | $1.71^{+0.11}_{-0.09}$ | $0.18^{+0.14}_{-0.12}$ |
| SGRB | 8–40 (Fermi) | $1.76^{+0.27}_{-0.21}$ | $0.78^{+0.02}_{-0.02}$ | $0.45^{+0.37}_{-0.44}$ | $1.60^{+0.18}_{-0.15}$ | $0.22^{+0.19}_{-0.16}$ |
| SGRB | 40–200 (Fermi) | $2.95^{+0.20}_{-0.14}$ | $0.03^{+0.03}_{-0.03}$ | $0.42^{+0.26}_{-0.38}$ | $1.56^{+0.07}_{-0.07}$ | $0.36^{+0.10}_{-0.15}$ |
| SGRB | 200–1000 (Fermi) | $1.87^{+0.16}_{-0.18}$ | $0.64^{+0.02}_{-0.02}$ | $0.26^{+0.36}_{-0.51}$ | $1.34^{+0.12}_{-0.11}$ | $0.24^{+0.16}_{-0.15}$ |

**Notes.**
[a] Two types of GRB, i.e., LGRB and SGRB.
[b] Logarithmic break frequency in the average PDS.

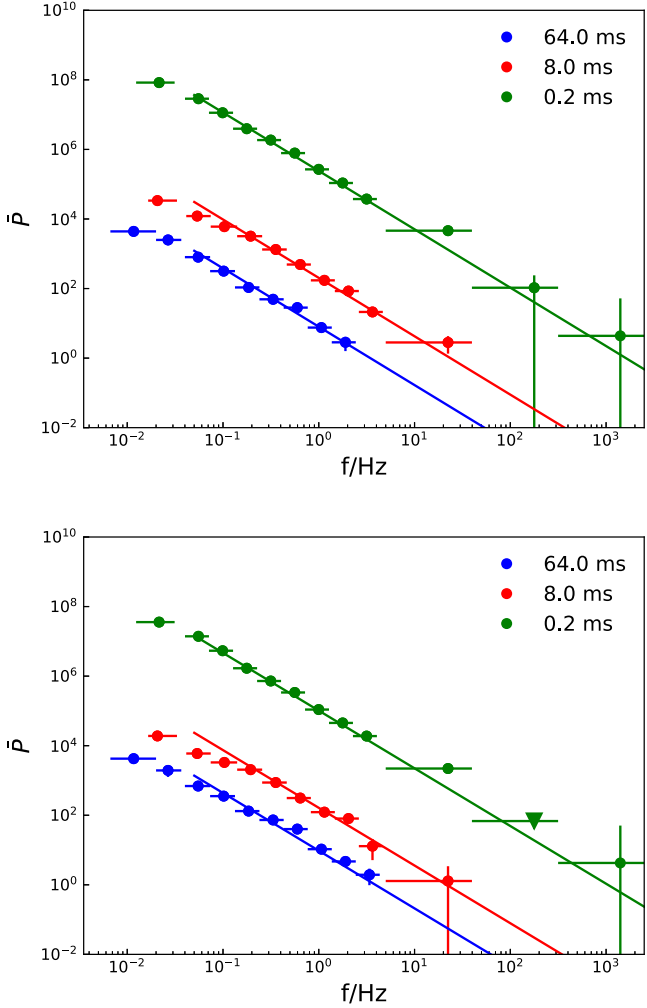

**Figure 3.** Average noise-subtracted PDSs of LGRBs with different time resolutions. The data and fitting results are from the 100 to 245 keV energy band (top) and from the 245 to 600 keV energy band (bottom). The blue, red, and green colors represent 64, 8, and 0.2 ms, respectively. The circles are the data points, and the triangles are upper limits of the data points. The best fittings with power-law indices of 1.58 and 1.37 are drawn with solid lines.

$2.84 \pm 1.12$ and $1.94 \pm 0.34$, respectively (as shown in Figure 7 and Table 4).

### 3.5. Energy Dependence

Figure 9 displays the relation between $\alpha_{\bar{P}}$ and its corresponding energy from Insight-HXMT, Swift/BAT, CGRO/BATSE, Fermi/GBM, and BeppoSAX/GRBM. For LGRBs (top panel), the fitting result shows $\alpha_{\bar{P}}$ as a function of energy with the relation

$$\alpha_{\bar{P}} = 2.59 \pm 0.14 \left(\frac{E}{1 \text{ keV}}\right)^{-0.09 \pm 0.01}. \qquad (11)$$

Our result is the same as the previous result of $\alpha_{\bar{P}} \propto E^{-0.09}$ in the frequency range $10^{-2}$–$10^0$ Hz (Beloborodov et al. 2000; Guidorzi et al. 2012; Dichiara et al. 2013). The Insight-HXMT data allow us to extend the energy of the $\alpha_{\bar{P}} - E$ relation to 2000 keV.

To compare the Insight-HXMT SGRBs with those from other satellites and reveal the trend of the spectral shape in different energy bands from soft to hard, we also analyze the SGRB average PDS of Fermi/GBM and Swift/BAT. For Swift/BAT, we divide the data into two energy bands (15–50 and 50–150 keV), and 13 bright SGRBs (GRBs 151221A, 060313, 061201, 080426, 090510, 101219A, 120804A, 130603B, 140930B, 151229A, 180204A, 191031D, and 201221D) are included in the sample. For Fermi/GBM, 12 bright SGRBs (bn190720710, 090802235, 110529034, 120817168, 140901821, 141011282, 160715298, 170708046, 171108656, 191031891, 191227723, and 201227635) are included in the sample, and their PDSs are calculated in three energy bands (8–40, 40–200, and 200–1000 keV). The results are also listed in Table 2. We obtain the value of $\alpha_{\bar{P}}$ for Swift/BAT: $1.86^{+0.16}_{-0.13}$ and $1.71^{+0.11}_{-0.09}$ in the energy bands 15–50 and 50–150 keV, respectively. For Fermi/GBM, the value of $\alpha_{\bar{P}}$ is $1.60^{+0.18}_{-0.15}$, $1.56^{+0.07}_{-0.07}$, and $1.34^{+0.12}_{-0.11}$ in the energy bands 8–40, 40–200, and 200–1000 keV, respectively. The results are

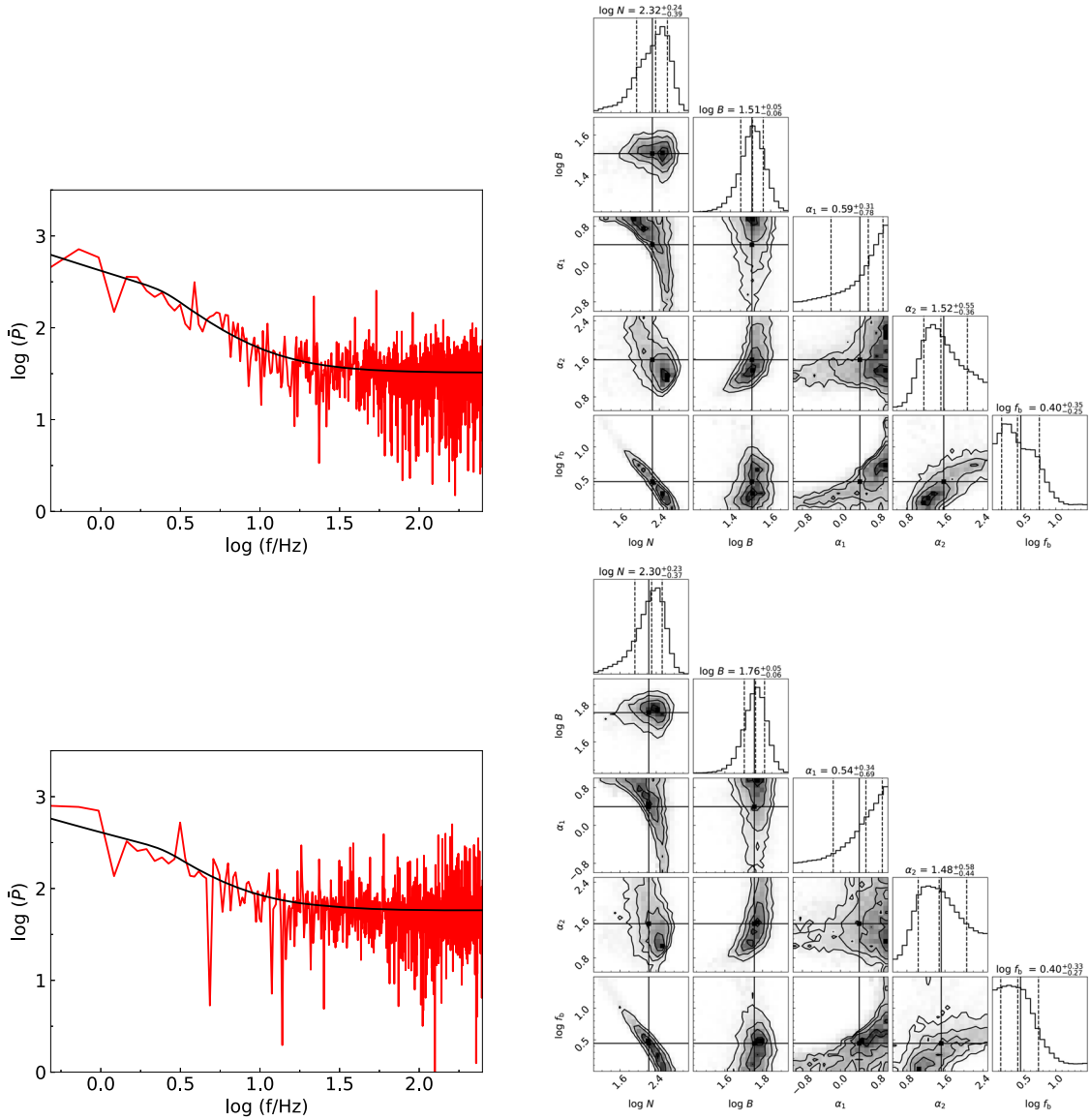

**Figure 4.** Left panels: fitting results of the average PDSs of SGRBs. Right panels: the distributions of parameters. The top and bottom panels correspond energies of 100–245 keV and 245–600 keV.

shown in the bottom panel of Figure 9, with the relation

$$\alpha_{\bar{P}} = 2.19 \pm 0.25 \left(\frac{E}{1 \text{ keV}}\right)^{-0.07 \pm 0.02}. \quad (12)$$

The index $\alpha_{\bar{P}}$ decreases from 1.86 to 1.34 for energy increasing from 8 to 1000 keV.

### 3.6. Dominant Timescale

For the BPL model, an individual PDS of GRBs presents an obvious break $f_b$ in the low-frequency region. $f_b$ is suggested to be linked to the dominant timescale $\tau$, which represents the timescale of the dominant pulse in a light curve, and is defined as $\tau = 1/(2\pi f_b)$.

The distribution of the dominant timescale $\tau$ for Insight-HXMT GRBs is shown in the side panel of Figure 10 and Table 4. For LGRBs, the typical value of $\tau$ for our sample is 1.58 s, and it varies from 0.18 to 12.95 s. The distribution is similar to the result from Swift (Guidorzi et al. 2016) ($\tau \sim 0.2$–30 s, with a typical value of 4.1 s, for the Swift sample), whose energy range is generally lower (15–150 keV). For SGRBs, $\tau$ ranges between 0.01 and 0.08 s, with a typical value of 0.02 s.

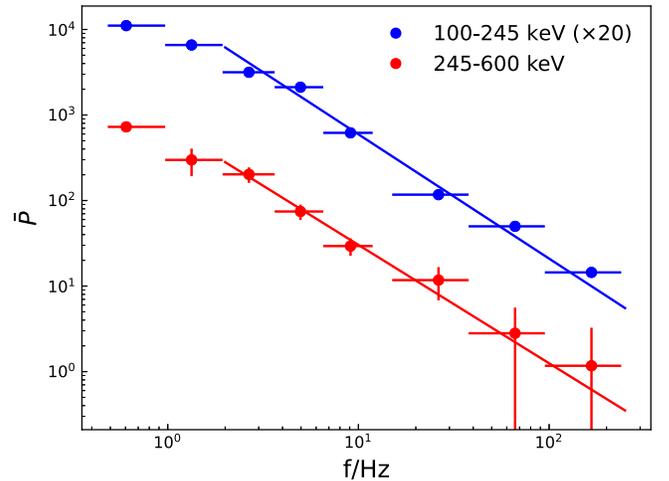

**Figure 5.** Average noise-subtracted PDSs of SGRBs with 2 ms time bin. The average PDSs at 100–245 and 245–600 keV are represented with blue and red colors, respectively. The solid lines are the best fittings with power-law indices of 1.52 for 100–245 keV and 1.48 for 245–600 keV.

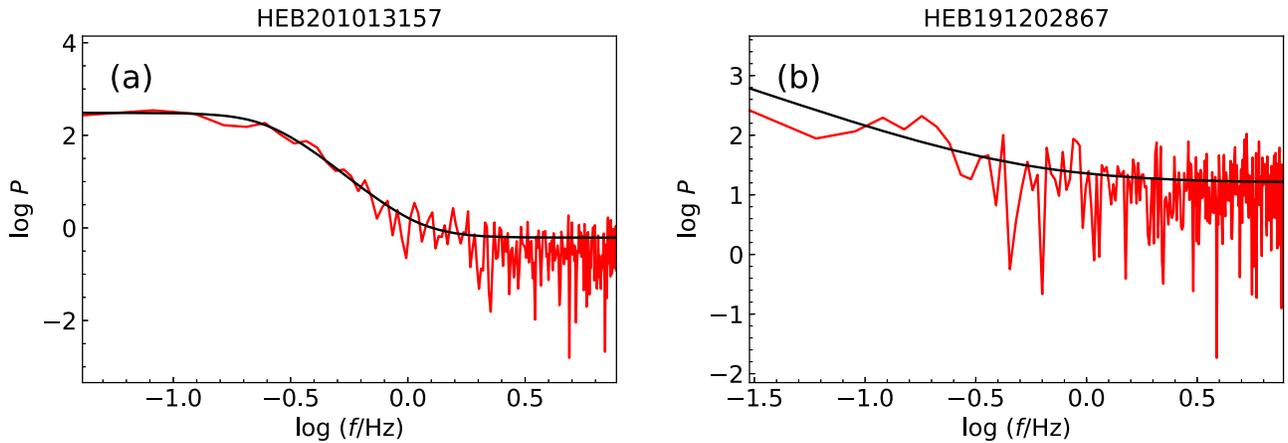

**Figure 6.** Examples of individual LGRB PDSs. The red and black lines represent the FFT PDS data and their corresponding best-fitting models. (a) The PDS of HEB201013157 with its best-fitting BPL model. (b) The PDS of HEB191202867 with its best-fitting PL model.

The dominant timescale $\tau$ is suggested to correspond to the pulse width of the light curve, which has a relation with the duration of the GRBs (Guidorzi et al. 2016). We also try to explore the $\tau$–$T_{90}$ relation with Insight-HXMT GRBs. As shown in Figure 10, we first fit the $\tau$–$T_{90}$ sets of Insight-HXMT LGRBs and find that they follow the relation

$$\tau = 10^{-0.90 \pm 0.28} \left(\frac{T_{90}}{1 \text{ s}}\right)^{0.86 \pm 0.21} \text{ s.} \quad (13)$$

The slopes of $\tau$–$T_{90}$ relations are similar to the results of Guidorzi et al. (2016) with the Swift LGRB sample, i.e., $0.78 \pm 0.13$.

On the other hand, $\tau$ and $T_{90}$ for SGRBs seem to also follow the relation estimated from the LGRB sample (red points in Figure 10). We try to consider $\tau$–$T_{90}$ of both LGRBs and SGRBs, and find that they are described by the relation

$$\tau = 10^{-0.92 \pm 0.11} \left(\frac{T_{90}}{1 \text{ s}}\right)^{0.86 \pm 0.09} \text{ s.} \quad (14)$$

This indicates that the dominant timescale is proportional to the duration of both LGRBs and SGRBs with a slope of $0.86 \pm 0.09$.

The ratio between $T_{90}$ and $\tau$, $T_{90}/\tau$, is relatively small for individual pulses, whereas it is larger for multipulse cases. This is exemplified by the light curve of HEB180828789 in Figure 11, where $T_{90}/\tau$ reaches a significant value of 20.56. The influence of factors such as the number of pulses and the interval between pulses can be observed from the light curve to affect the ratio of $T_{90}$ and $\tau$. Guidorzi et al. (2016) also explained why there is no apparent correlation between $\tau$ and $T_{90}$ in the Swift sample. $T_{90}/\tau$ exhibits distinct variations among different GRBs. These results exclude the possibility of $\tau$ and $T_{90}$ having a common origin.

There are three zones outside the $T_{90}$–$\tau$ area, i.e., the zone where the timescale $\tau$ of GRBs is close to the duration $T_{90}$ (Zone I), the zone of larger $\tau$ and smaller $T_{90}$ (Zone II), and the zone of smaller $\tau$ and larger $T_{90}$ (Zone III). Both $\tau$ and $T_{90}$ in a single-pulse burst are mainly contributed by the same monopulse. These suggest that the single-pulse GRBs are supposed to be located in Zone I. To verify this point, we constructed 1000 light curves of single-pulse GRBs (750 LGRBs and 250 SGRBs), and calculated their $\tau$ and $T_{90}$.[9]

As shown in Figure 10, 7/8 single-pulse GRBs (empty circles) in our sample are located in the zone from the simulation results (Zone I). Five LGRBs (HEB190806675, HEB190605110, HEB180404091, HEB181201111, HEB200227305) and two SGRBs (HEB190610477, HEB190906045) are located in Zone I. For LGRBs (SGRBs), the average value of $T_{90}/\tau$ is ~4.80 (~3.20), which is close to the simulation results, i.e., 5.52 (2.68). Since the timescale $\tau$ (approximate FWHM of the pulse) is shorter than $T_{90}$ (the time for 90% of a light curve), it is reasonable to suppose that there are no GRBs in Zone II. There are no GRBs are located in Zone III, which suggests that long-lasting GRBs with a short dominant timescale ($T_{90} \gg \tau$) are unlikely to occur (Guidorzi et al. 2016).

### 3.7. Periodic Signal

The individual PDS of a GRB provides an important approach to searching for periodic and quasi-periodic signals from GRB prompt emission (Kruger et al. 2002; Castro-Tirado et al. 2021). We perform a systematic investigation of periodic and quasi-periodic signals based on the Insight-HXMT GRBs. No signal above the $3\sigma$ significance threshold is found using the average PDS and individual PDSs. The most significant case in our sample is HEB200412381 with $1.74\sigma$ significance.

### 4. Discussion

The power-law index was found to be compatible with $5/3$, which is consistent with the prediction from the Kolmogorov spectrum (Beloborodov et al. 2000). Panaitescu et al. (1999) simulated the light curves arising from the internal shock

---

[9] We produced pulses for single-pulse GRBs with the empirical pulse model (Kocevski et al. 2003)

$$C(t) = C_p \left(\frac{t}{t_m}\right)^r \left[\frac{d}{d+r} + \frac{r}{d+r}\left(\frac{t}{t_m}\right)^{r+1}\right]^{-(r+d)/(r+1)}, \quad (15)$$

where $t_m$ is the time of peak net counts $C_p$, and $r$ and $d$ are the power-law rising and decaying indices, respectively. We generate a sample of 750 LGRBs and 250 SGRBs following normal distributions based on $\log(t_m/\text{s}) = 0.61 \pm 0.74$, $r = 1.48 \pm 0.61$, and $d = 2.44 \pm 0.76$ for LGRBs (Kocevski et al. 2003), and $\log(t_m/\text{s}) = -0.25 \pm 0.49$, $r = 88.54 \pm 113.17$, and $d = 12.64 \pm 9.28$ for SGRBs (Li et al. 2022).

Table 3
PDS Fitting Results of 64 LGRBs

| Name | $\log N$ | $\log B$ | $\alpha$ | $\log(f_b/\mathrm{Hz})$ | $\log \mathrm{BF}^a$ | Best Model[b] | $p_{\mathrm{KS}}$ | Work Mode[c] |
|---|---|---|---|---|---|---|---|---|
| HEB170626400 | $1.89^{+0.28}_{-0.53}$ | $0.27^{+0.04}_{-0.05}$ | $3.52^{+0.97}_{-0.98}$ | $-0.47^{+0.32}_{-0.15}$ | 1.01 | BPL | 0.54 | Normal |
| HEB170726793 | $1.31^{+0.11}_{-0.14}$ | $0.66^{+0.07}_{-0.05}$ | $2.78^{+0.87}_{-1.09}$ | $-0.14^{+0.17}_{-0.10}$ | 0.11 | BPL | 0.64 | Normal |
| HEB170817908 | $0.61^{+0.13}_{-0.13}$ | $0.13^{+0.02}_{-0.02}$ | $1.81^{+0.30}_{-0.35}$ | ... | $-0.13$ | PL | 0.46 | Normal |
| HEB170826818 | $1.89^{+0.24}_{-0.37}$ | $0.47^{+0.06}_{-0.05}$ | $1.90^{+0.38}_{-0.52}$ | $-0.50^{+0.39}_{-0.26}$ | 0.31 | BPL | 0.54 | Normal |
| HEB170906029 | $0.93^{+0.13}_{-0.11}$ | $0.70^{+0.08}_{-0.07}$ | $1.23^{+0.23}_{-0.28}$ | ... | $-2.02$ | PL | 0.72 | GRB |
| HEB170923188 | $2.44^{+0.27}_{-0.29}$ | $0.88^{+0.09}_{-0.08}$ | $1.31^{+0.23}_{-0.38}$ | $-1.10^{+0.29}_{-0.34}$ | 0.45 | BPL | 0.76 | Normal |
| HEB170926340 | $0.18^{+0.61}_{-0.29}$ | $0.14^{+0.02}_{-0.02}$ | $1.85^{+0.59}_{-1.08}$ | ... | $-0.66$ | PL | 0.55 | Normal |
| HEB171210492 | $2.87^{+0.48}_{-0.67}$ | $1.32^{+0.03}_{-0.03}$ | $3.18^{+0.88}_{-1.09}$ | $-1.33^{+0.33}_{-0.27}$ | 0.31 | BPL | 0.65 | Normal |
| HEB171215705 | $1.06^{+0.15}_{-0.13}$ | $1.02^{+0.08}_{-0.06}$ | $1.13^{+0.20}_{-0.24}$ | ... | $-0.21$ | PL | 0.35 | Normal |
| HEB180103047 | $2.89^{+0.24}_{-0.36}$ | $0.62^{+0.03}_{-0.03}$ | $2.36^{+0.33}_{-0.47}$ | $-0.97^{+0.27}_{-0.20}$ | $-0.95$ | BPL | 0.07 | Normal |
| HEB180111695 | $2.19^{+0.23}_{-0.32}$ | $1.04^{+0.07}_{-0.06}$ | $1.50^{+0.34}_{-0.57}$ | $-0.62^{+0.45}_{-0.30}$ | 0.31 | BPL | 0.32 | GRB |
| HEB180202211 | $-0.81^{+0.44}_{-1.63}$ | $0.67^{+0.07}_{-0.06}$ | $3.07^{+0.23}_{-0.51}$ | ... | $-0.35$ | PL | 0.37 | Normal |
| HEB180210517 | $2.71^{+0.23}_{-0.23}$ | $1.24^{+0.07}_{-0.05}$ | $0.96^{+0.14}_{-0.18}$ | $-1.55^{+0.27}_{-0.36}$ | 0.47 | BPL | 0.35 | Normal |
| HEB180221520 | $1.12^{+0.12}_{-0.11}$ | $1.21^{+0.04}_{-0.04}$ | $1.27^{+0.17}_{-0.19}$ | ... | $-0.23$ | PL | 0.26 | Normal |
| HEB180305393 | $0.12^{+0.09}_{-0.09}$ | $0.01^{+0.00}_{-0.00}$ | $2.26^{+0.18}_{-0.20}$ | ... | $-0.16$ | PL | 0.54 | Normal |
| HEB180326143 | $2.04^{+0.19}_{-0.26}$ | $0.42^{+0.04}_{-0.04}$ | $2.01^{+0.30}_{-0.36}$ | $-0.56^{+0.24}_{-0.17}$ | 0.94 | BPL | 0.45 | Normal |
| HEB180404091 | $3.45^{+0.35}_{-0.45}$ | $1.54^{+0.02}_{-0.02}$ | $2.78^{+0.65}_{-1.08}$ | $-1.50^{+0.27}_{-0.25}$ | 0.43 | BPL | 0.24 | Normal |
| HEB180411359 | $3.74^{+0.20}_{-0.29}$ | $2.18^{+0.01}_{-0.01}$ | $2.34^{+0.45}_{-0.65}$ | $-1.41^{+0.27}_{-0.19}$ | 0.02 | BPL | 0.21 | Normal |
| HEB180718762 | $3.42^{+0.34}_{-0.34}$ | $1.02^{+0.05}_{-0.04}$ | $1.43^{+0.17}_{-0.22}$ | $-1.71^{+0.24}_{-0.32}$ | 0.26 | BPL | 0.65 | Normal |
| HEB180804554 | $0.94^{+0.24}_{-0.19}$ | $1.39^{+0.04}_{-0.03}$ | $1.26^{+0.21}_{-0.26}$ | ... | $-0.50$ | PL | 0.59 | Normal |
| HEB180804930 | $1.92^{+0.27}_{-0.37}$ | $0.58^{+0.06}_{-0.06}$ | $2.76^{+0.77}_{-1.27}$ | $-0.42^{+0.29}_{-0.18}$ | 0.30 | BPL | 0.43 | Normal |
| HEB180828789 | $1.87^{+0.18}_{-0.26}$ | $0.13^{+0.02}_{-0.03}$ | $2.69^{+0.37}_{-0.49}$ | $-0.38^{+0.19}_{-0.14}$ | 1.65 | BPL | 0.55 | Normal |
| HEB180925609 | $1.11^{+0.09}_{-0.08}$ | $0.68^{+0.09}_{-0.08}$ | $0.96^{+0.14}_{-0.16}$ | ... | $-0.07$ | PL | 0.53 | Normal |
| HEB181014479 | $-2.28^{+1.72}_{-1.66}$ | $1.56^{+0.02}_{-0.02}$ | $3.21^{+1.05}_{-1.12}$ | ... | 0.61 | PL | 0.31 | Normal |
| HEB181201111 | $3.81^{+0.24}_{-0.34}$ | $0.53^{+0.01}_{-0.01}$ | $2.82^{+0.21}_{-0.23}$ | $-1.49^{+0.16}_{-0.13}$ | 0.21 | BPL | 0.20 | Normal |
| HEB181212692 | $2.20^{+0.35}_{-0.56}$ | $0.08^{+0.01}_{-0.01}$ | $3.41^{+0.66}_{-0.87}$ | $-0.64^{+0.29}_{-0.18}$ | 0.15 | BPL | 0.48 | Normal |
| HEB190103877 | $2.88^{+0.31}_{-0.42}$ | $1.27^{+0.03}_{-0.03}$ | $2.93^{+0.77}_{-1.14}$ | $-0.98^{+0.30}_{-0.20}$ | 0.63 | BPL | 0.49 | Normal |
| HEB190110725 | $-1.05^{+1.06}_{-0.80}$ | $0.45^{+0.07}_{-0.06}$ | $3.37^{+1.39}_{-1.13}$ | ... | $-0.55$ | PL | 0.49 | Normal |
| HEB190117608 | $1.67^{+0.10}_{-0.11}$ | $0.26^{+0.03}_{-0.03}$ | $3.35^{+0.49}_{-0.59}$ | $-0.14^{+0.09}_{-0.07}$ | 0.84 | BPL | 0.43 | Normal |
| HEB190131964 | $3.05^{+0.27}_{-0.28}$ | $1.36^{+0.06}_{-0.05}$ | $1.18^{+0.17}_{-0.22}$ | $-1.46^{+0.27}_{-0.34}$ | 0.39 | BPL | 0.30 | Normal |
| HEB190215771 | $0.11^{+0.32}_{-0.26}$ | $0.85^{+0.03}_{-0.03}$ | $2.00^{+0.30}_{-0.37}$ | ... | $-0.37$ | PL | 0.44 | Normal |
| HEB190321931 | $2.06^{+0.69}_{-1.05}$ | $-0.04^{+0.01}_{-0.01}$ | $2.66^{+0.72}_{-1.12}$ | $-0.65^{+0.49}_{-0.40}$ | 0.32 | BPL | 0.53 | Normal |
| HEB190324348 | $3.91^{+0.30}_{-0.34}$ | $1.44^{+0.02}_{-0.02}$ | $1.94^{+0.21}_{-0.29}$ | $-1.73^{+0.23}_{-0.24}$ | 0.16 | BPL | 0.17 | Normal |
| HEB190326313 | $2.84^{+0.17}_{-0.24}$ | $1.51^{+0.03}_{-0.02}$ | $1.65^{+0.27}_{-0.36}$ | $-0.97^{+0.28}_{-0.20}$ | 0.54 | BPL | 0.37 | Normal |
| HEB190531840 | $2.81^{+0.11}_{-0.14}$ | $0.25^{+0.03}_{-0.03}$ | $2.19^{+0.12}_{-0.13}$ | $-0.75^{+0.10}_{-0.08}$ | 0.53 | BPL | 0.57 | Normal |
| HEB190604446 | $0.25^{+0.98}_{-0.39}$ | $0.86^{+0.07}_{-0.06}$ | $1.56^{+0.50}_{-1.09}$ | ... | $-0.29$ | PL | 0.53 | Normal |
| HEB190605110 | $2.47^{+0.35}_{-0.36}$ | $0.98^{+0.10}_{-0.09}$ | $1.03^{+0.19}_{-0.25}$ | $-1.41^{+0.38}_{-0.51}$ | 0.13 | BPL | 0.44 | Normal |
| HEB190613449 | $0.26^{+0.27}_{-0.20}$ | $0.32^{+0.05}_{-0.05}$ | $1.64^{+0.38}_{-0.51}$ | ... | $-0.57$ | PL | 0.40 | Normal |
| HEB190615612 | $0.72^{+0.52}_{-1.04}$ | $0.40^{+0.05}_{-0.05}$ | $2.90^{+1.56}_{-1.46}$ | $-0.80^{+0.46}_{-0.70}$ | 0.39 | BPL | 0.52 | Normal |
| HEB190620507 | $4.26^{+0.49}_{-0.48}$ | $0.83^{+0.02}_{-0.02}$ | $2.17^{+0.19}_{-0.24}$ | $-1.91^{+0.27}_{-0.27}$ | 0.02 | BPL | 0.78 | Normal |
| HEB190806675 | $2.58^{+0.44}_{-0.80}$ | $1.32^{+0.04}_{-0.03}$ | $1.76^{+0.50}_{-1.23}$ | $-1.86^{+0.80}_{-0.55}$ | 0.28 | BPL | 0.51 | Normal |
| HEB190828783 | $0.66^{+0.16}_{-0.14}$ | $0.48^{+0.07}_{-0.06}$ | $1.64^{+0.29}_{-0.38}$ | ... | $-0.02$ | PL | 0.43 | Normal |
| HEB190906767 | $0.91^{+0.09}_{-0.09}$ | $0.29^{+0.04}_{-0.04}$ | $1.80^{+0.22}_{-0.24}$ | ... | $-0.07$ | PL | 0.64 | Normal |
| HEB190928551 | $2.94^{+0.18}_{-0.23}$ | $-0.24^{+0.02}_{-0.02}$ | $2.83^{+0.21}_{-0.24}$ | $-0.93^{+0.13}_{-0.11}$ | 1.17 | BPL | 0.38 | Normal |
| HEB191019970 | $2.74^{+0.56}_{-0.99}$ | $1.19^{+0.03}_{-0.03}$ | $3.00^{+0.92}_{-1.23}$ | $-1.27^{+0.55}_{-0.32}$ | 0.22 | BPL | 0.46 | Normal |
| HEB191019971 | $3.05^{+0.21}_{-0.29}$ | $1.36^{+0.02}_{-0.02}$ | $4.23^{+0.98}_{-0.57}$ | $-1.07^{+0.14}_{-0.09}$ | 0.18 | BPL | 0.08 | Normal |
| HEB191202867 | $-1.44^{+2.37}_{-1.39}$ | $0.58^{+0.07}_{-0.07}$ | $2.28^{+1.32}_{-1.59}$ | ... | $-2.32$ | PL | 0.71 | Normal |
| HEB191218112 | $4.15^{+0.30}_{-0.39}$ | $1.39^{+0.01}_{-0.01}$ | $2.36^{+0.21}_{-0.26}$ | $-1.63^{+0.22}_{-0.17}$ | 0.58 | BPL | 0.36 | Normal |
| HEB191221860 | $0.84^{+0.06}_{-0.06}$ | $0.28^{+0.05}_{-0.05}$ | $1.47^{+0.11}_{-0.12}$ | ... | $-0.31$ | PL | 0.41 | Normal |
| HEB191227069 | $2.59^{+0.17}_{-0.22}$ | $-0.17^{+0.02}_{-0.02}$ | $2.09^{+0.11}_{-0.12}$ | $-0.81^{+0.15}_{-0.12}$ | 0.43 | BPL | 0.45 | Normal |
| HEB200111632 | $-0.57^{+0.62}_{-0.53}$ | $0.31^{+0.05}_{-0.05}$ | $3.52^{+1.05}_{-0.95}$ | ... | $-0.30$ | PL | 0.23 | Normal |
| HEB200125863 | $1.41^{+0.21}_{-0.28}$ | $-0.42^{+0.05}_{-0.05}$ | $2.86^{+0.56}_{-0.76}$ | $-0.06^{+0.23}_{-0.16}$ | 0.90 | BPL | 0.51 | Normal |
| HEB200219998 | $-0.14^{+0.55}_{-0.44}$ | $0.75^{+0.05}_{-0.06}$ | $2.65^{+0.88}_{-1.08}$ | ... | $-0.28$ | PL | 0.61 | Normal |
| HEB200227305 | $2.23^{+0.98}_{-1.51}$ | $1.07^{+0.04}_{-0.04}$ | $3.27^{+1.26}_{-1.13}$ | $-1.65^{+0.51}_{-0.47}$ | 0.22 | BPL | 0.46 | Normal |
| HEB200412381 | $1.92^{+0.17}_{-0.20}$ | $-0.02^{+0.00}_{-0.00}$ | $2.96^{+0.40}_{-0.52}$ | $-0.28^{+0.15}_{-0.12}$ | 1.79 | BPL | 0.72 | Normal |
| HEB200413712 | $-0.37^{+1.37}_{-0.62}$ | $0.50^{+0.06}_{-0.07}$ | $2.71^{+1.15}_{-1.35}$ | ... | $-0.04$ | PL | 0.48 | Normal |
| HEB200619108 | $2.52^{+0.28}_{-0.40}$ | $0.93^{+0.04}_{-0.04}$ | $2.20^{+0.49}_{-0.93}$ | $-0.83^{+0.35}_{-0.25}$ | 0.04 | BPL | 0.51 | Normal |
| HEB200716956 | $1.09^{+0.15}_{-0.18}$ | $-0.63^{+0.09}_{-0.08}$ | $3.84^{+0.80}_{-0.75}$ | $0.16^{+0.11}_{-0.08}$ | 1.61 | BPL | 0.72 | Normal |

Table 3
(Continued)

| Name | log N | log B | α | log($f_b$/Hz) | log BF[a] | Best Model[b] | $p_{KS}$ | Work Mode[c] |
|---|---|---|---|---|---|---|---|---|
| HEB201013157 | $2.49^{+0.20}_{-0.25}$ | $-0.28^{+0.03}_{-0.03}$ | $3.95^{+0.44}_{-0.49}$ | $-0.60^{+0.11}_{-0.09}$ | 2.15 | BPL | 0.22 | Normal |
| HEB201105229 | $-2.12^{+0.71}_{-0.84}$ | $0.37^{+0.05}_{-0.05}$ | $3.95^{+1.06}_{-0.75}$ | ... | $-0.09$ | PL | 0.38 | GRB |
| HEB210112068 | $0.86^{+0.11}_{-0.09}$ | $0.81^{+0.05}_{-0.05}$ | $1.16^{+0.17}_{-0.20}$ | ... | $-0.01$ | PL | 0.46 | Normal |
| HEB210121779 | $3.17^{+0.23}_{-0.32}$ | $0.71^{+0.03}_{-0.02}$ | $1.66^{+0.13}_{-0.16}$ | $-1.35^{+0.28}_{-0.21}$ | 0.20 | BPL | 0.28 | GRB |
| HEB210207911 | $-2.19^{+1.87}_{-1.62}$ | $0.75^{+0.06}_{-0.07}$ | $2.31^{+1.51}_{-1.79}$ | ... | $-0.13$ | PL | 0.78 | GRB |
| HEB210213286 | $1.55^{+0.05}_{-0.04}$ | $1.51^{+0.02}_{-0.02}$ | $1.35^{+0.06}_{-0.07}$ | ... | $-0.37$ | PL | 0.39 | Normal |

**Notes.**
[a] Obtained from Equation (10), whose positive or negative sign depends on whether the best-fitting model is the BPL or PL model.
[b] The p-value of the K-S test, for a null hypothesis that the individual PDS is consistent with the $\chi^2$ distribution.
[c] Normal mode or GRB mode. The Normal mode corresponds the energy range 100–600 keV, while the GRB mode corresponds the range 400–2000 keV.

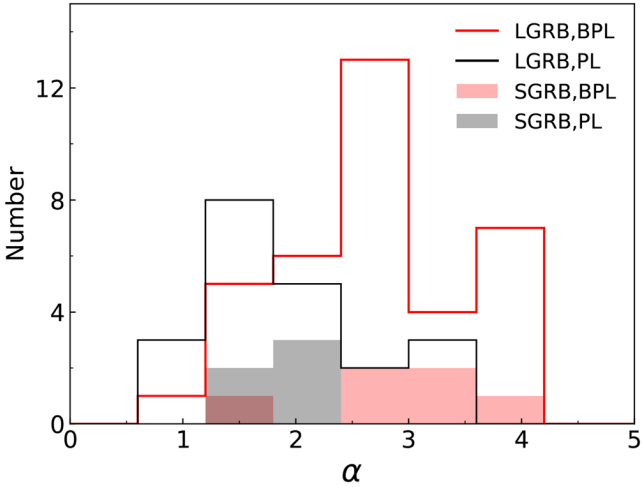

**Figure 7.** The distributions of the individual PDS index α, which are represented by the solid lines for LGRBs and filled lines for SGRBs. The distributions of α for the PL and BPL models are marked with black and red colors, respectively.

model. The PDSs of the simulated light curves exhibit a ∼5/3 index up to 1 Hz. Zhang & Zhang (2014) simulated the light curves within the framework of the internal-collision-induced magnetic reconnection and turbulence (ICMART) model (Zhang & Yan 2011), and the corresponding PDSs can exhibit indices of 1−3, which are dependent on different tunable parameters. It is suggested that the index of the observational average PDS of LGRBs varies from ∼1.5 to ∼2, which depends on the energy passband (Beloborodov et al. 2000; Dichiara et al. 2013). Our result indicates that the average PDS index of SGRBs also undergoes the steep-to-shallow evolution from soft to hard energy bands. This behavior may result from the relation between energy and pulse width (Fenimore et al. 1995; Beloborodov et al. 2000; Dichiara et al. 2013). The physical origin of this relation is unclear, e.g., it may be due to the conversion of bulk motion into gamma rays (Fenimore et al. 1995). The similarity between LGRBs and SGRBs in terms of the energy dependence indicates that the two types of GRB may have similar physical origins. The typical values of α for the individual PDSs of Insight-HXMT LGRBs are 2.49 ± 0.78 and 1.95 ± 0.74 in the group with a dominant timescale (BPL model) and the group with no dominant timescale (PL model),

Table 4
Fitting Results of the Distribution of Individual PDS Parameters

| Type[a] | Fitting Model | α[b] | τ[c] (s) |
|---|---|---|---|
| LGRB | BPL | 2.49 ± 0.78 | $1.58^{+3.66}_{-1.11}$ |
| LGRB | PL | 1.95 ± 0.74 | ... |
| SGRB | BPL | 2.84 ± 1.12 | $0.02^{+0.03}_{-0.01}$ |
| SGRB | PL | 1.94 ± 0.34 | ... |

**Notes.**
[a] Two types of GRB, i.e., LGRB and SGRB.
[b] The mean value of the power-law index of the individual PDS.
[c] The mean value of the dominant timescale, which is derived from $\tau = 1/2\pi f_b$, where $f_b$ is the break frequency of an individual PDS.

respectively. The results are compatible with the mean values of α in the Swift sample: 2.9 for the BPL sample and 2.1 for the PL sample (Guidorzi et al. 2016). There is an obvious difference between the groups with and without a dominant timescale. This may be because the GRB light curves are the results of superposing a number of pulses with different timescales (Dichiara et al. 2016).

The results show that the typical values of α of the groups with and without a dominant timescale for SGRBs are 2.84 ± 1.12 and 1.94 ± 0.34, which is similar to the α values of LGRBs. This may be caused by the similar spectral slopes for LGRBs and SGRBs. Assuming that a GRB prompt emission light curve consists of a number of pulses, the normalized PDS of the $i$th pulse in the light curve can be given as (Belli 1992; Lazzati 2002; Guidorzi et al. 2016)

$$P_i = \frac{C_i^2}{C_{max}^2} \frac{\tau_i^{\alpha_i}}{1 + (2\pi f \tau_i)^{\alpha_i}}, \quad (16)$$

where $\tau_i$ is the duration, $\alpha_i$ is the slope, $C_i$ is the peak counts, while $C_{max}$ is the maximum peak counts of pulses in the GRB light curve. One can see that the total PDS is dominated by some specific $\tau_i$ (dominant timescale) with high $C_i$ in the light curve (as shown in Figure 12), which was first proposed by Dichiara et al. (2016). The light curve of GRBs for this case should be included in the BPL group (Dichiara et al. 2016).

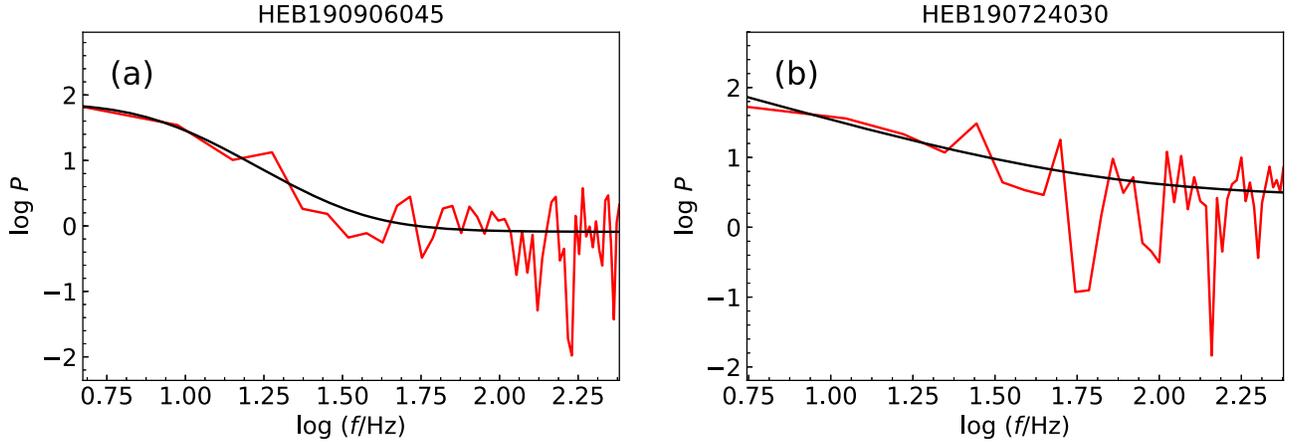

**Figure 8.** Examples of individual SGRB PDSs. The red and black lines represent the FFT PDS data and their corresponding best-fitting model. (a) The PDS of HEB190906045 with its best-fitting BPL model. (b) The PDS of HEB190724030 with its best-fitting PL model.

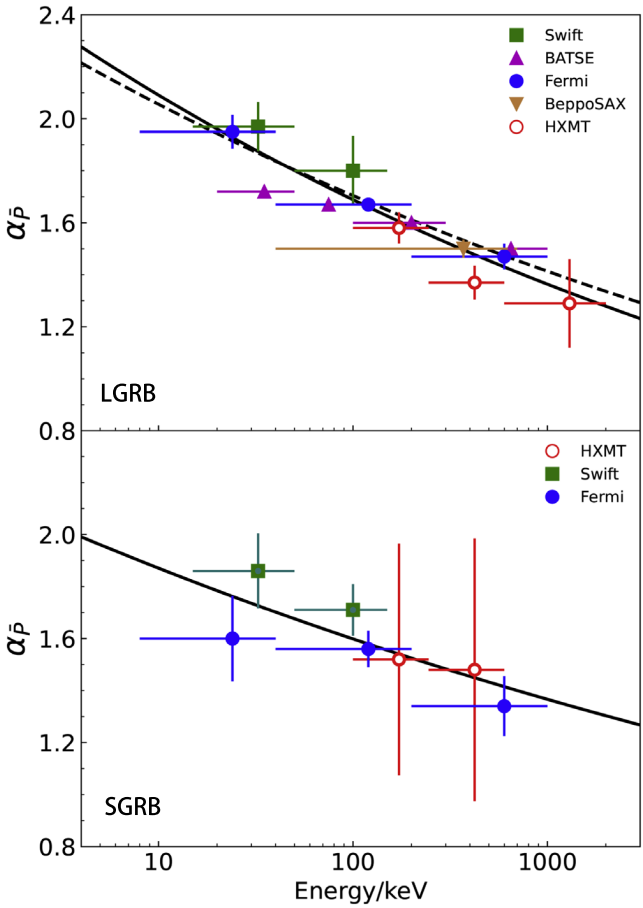

**Figure 9.** The relation between $\alpha_{\bar{P}}$ and energy $E$, for LGRBs (top panel) and SGRBs (bottom panel). The LGRB $\alpha_{\bar{P}}$ of Swift and BATSE are presented from Guidorzi et al. (2012) and Beloborodov et al. (2000), respectively. The LGRB $\alpha_{\bar{P}}$ of Fermi and BeppoSAX are presented from Dichiara et al. (2013). The dashed line represents the best-fitting average PDS from previous work. The solid line represents the best-fitting PDS after including Insight-HXMT results.

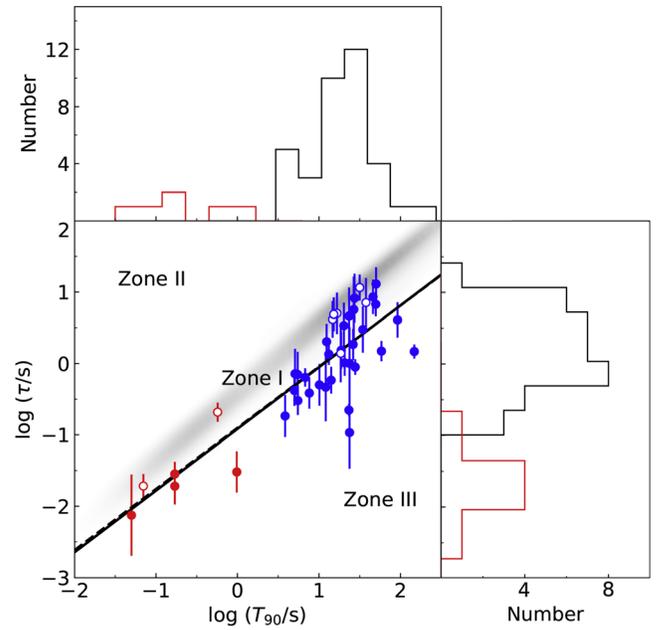

**Figure 10.** Relation between $\tau$ and $T_{90}$. The blue solid circles are data points of LGRBs, the red solid circles are data points of SGRBs. The solid line is the best linear fit of all GRBs, and the dashed line is the best linear fit of LGRBs. The single-pulse GRBs are marked by the empty circles. The gray shaded area is the simulated region where single-pulse cases should be located theoretically. The side panels show the histograms of $\log T_{90}$ and $\log \tau$, where black lines and red lines represent LGRBs and SGRBs, respectively.

and $\tau_i$ have a power-law relation,

$$C_i/C_{\max} \sim \tau_i^k, \tag{17}$$

then Equation (16) can be written as

$$P_i \sim \frac{\tau_i^{\alpha_i+2k}}{1+(2\pi f \tau_i)^{\alpha_i}}. \tag{18}$$

The PDS of the $i$th pulse at the break frequency $P_{b,i}$ is

$$P_{b,i}(f_{b,i}) \sim \tau_i^{\alpha_i+2k} = f_{b,i}^{-(\alpha_i+2k)}. \tag{19}$$

One can see that, if $k > 0$, the pulse with a higher break frequency exhibits a lower power density, which means that it has a lower weight in the total PDS of GRBs, and the total PDS still shows a break in the lower frequency range (as shown in

The GRBs in the PL group arise because their pulses have different durations and similar weights in the total variance (Dichiara et al. 2016). As shown in the bottom panel of Figure 13, for a GRB in the PL group, the break power density of the $i$th pulse $P_{b,i} \sim (C_i/C_{\max})^2(\tau_i^{\alpha_i}/2)$ may decay if its break frequency $f_{b,i}$ ($1/(2\pi\tau_i)$) increases. Here, we assume $C_i/C_{\max}$

Table 5
PDS Fitting Results of 11 SGRBs

| Name | log N | log B | α | log($f_b$/Hz) | log BF[a] | Best Model[b] | $p_{KS}$ | Work Mode[c] |
|---|---|---|---|---|---|---|---|---|
| HEB170801208 | $3.13^{+0.98}_{-0.61}$ | $0.22^{+0.03}_{-0.03}$ | $2.30^{+0.69}_{-0.65}$ | ... | −0.28 | PL | 0.44 | Normal |
| HEB170802637 | $2.73^{+0.13}_{-0.15}$ | $1.12^{+0.02}_{-0.02}$ | $4.24^{+0.78}_{-0.54}$ | $0.72^{+0.07}_{-0.06}$ | 3.52 | BPL | 0.98 | Normal |
| HEB171030728 | $1.06^{+0.22}_{-0.21}$ | $0.08^{+0.01}_{-0.01}$ | $1.15^{+0.42}_{-0.64}$ | $1.76^{+0.32}_{-0.45}$ | 0.51 | BPL | 0.71 | Normal |
| HEB171223818 | $2.09^{+0.21}_{-0.25}$ | $0.21^{+0.03}_{-0.03}$ | $3.25^{+0.59}_{-0.85}$ | $0.99^{+0.14}_{-0.13}$ | 0.95 | BPL | 0.52 | Normal |
| HEB180402406 | $2.36^{+0.21}_{-0.22}$ | $0.80^{+0.05}_{-0.05}$ | $3.05^{+0.60}_{-0.86}$ | $0.90^{+0.10}_{-0.13}$ | 0.49 | BPL | 0.25 | Normal |
| HEB180618030 | $2.62^{+0.27}_{-0.32}$ | $1.39^{+0.03}_{-0.03}$ | $1.59^{+0.41}_{-0.69}$ | ... | −0.75 | PL | 0.36 | Normal |
| HEB190326316 | $2.71^{+1.03}_{-0.83}$ | $-0.38^{+0.06}_{-0.06}$ | $1.47^{+0.52}_{-0.47}$ | ... | −0.02 | PL | 0.73 | Normal |
| HEB190610477 | $2.86^{+0.12}_{-0.12}$ | $0.77^{+0.04}_{-0.04}$ | $1.58^{+0.13}_{-0.15}$ | $0.30^{+0.04}_{-0.08}$ | 0.17 | BPL | 0.96 | Normal |
| HEB190724030 | $3.50^{+0.54}_{-0.35}$ | $0.34^{+0.06}_{-0.05}$ | $2.18^{+0.39}_{-0.31}$ | ... | −0.31 | PL | 0.71 | Normal |
| HEB190903721 | $3.29^{+0.34}_{-0.36}$ | $1.23^{+0.03}_{-0.03}$ | $2.14^{+0.45}_{-0.51}$ | ... | −0.38 | PL | 0.09 | Normal |
| HEB190906045 | $1.27^{+0.19}_{-0.20}$ | $-0.09^{+0.01}_{-0.01}$ | $3.79^{+0.91}_{-0.83}$ | $1.21^{+0.05}_{-0.10}$ | 0.63 | BPL | 0.95 | Normal |

**Notes.**
[a] Obtained from Equation (10), whose positive or negative sign depends on whether the best-fitting model is the BPL or PL model.
[b] The p-value of the K-S test, for a null hypothesis that the individual PDS is consistent with the $\chi^2$ distribution.
[c] Normal mode or GRB mode. The Normal mode corresponds the energy range 100–600 keV, while the GRB mode corresponds the range 400–2000 keV.

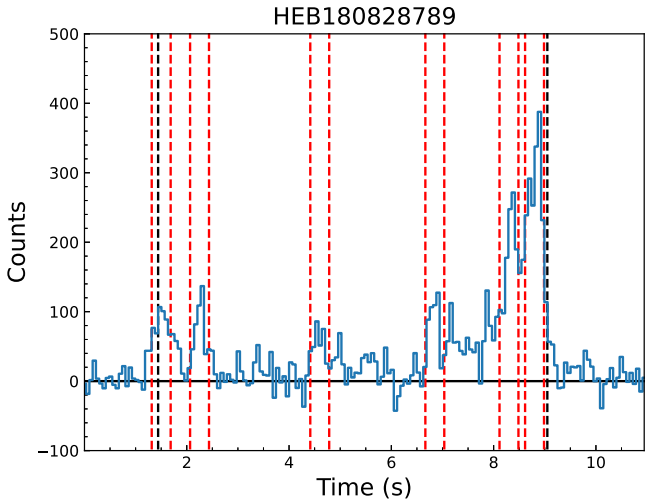

**Figure 11.** Example of a multipulse Insight-HXMT GRB, HEB180828789, with dominant timescale $\tau$. The solid line is the light curve, the black dashed lines are the boundaries of $T_{90}$, and the red dashed lines are the boundaries of $\tau$. The PDS of HEB180828789 is fit with the BPL model with $\tau = 0.37$ s, while $T_{90} = 7.61$, which is 20.56 times $\tau$.

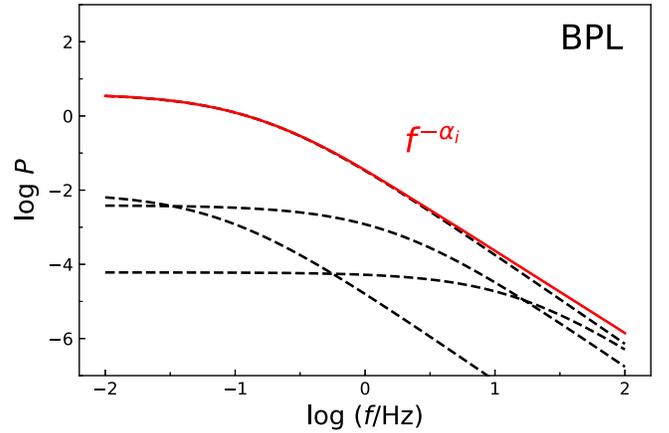

**Figure 12.** The total PDS is contributed by the PDSs of different pulses with specific $\tau_i$. The PDSs of different pulses $P_i(f)$ are represented with the black dashed lines, and the total PDS is represented with the red solid line. (The sketch of this case was proposed by Dichiara et al. 2016.)

the top panel of Figure 13). If $k < -\alpha_i/2$, the pulse with the highest break frequency dominates the total PDS with more weight, and the total PDS will break in a relatively higher frequency range (as shown in the middle panel of Figure 13). If $-\alpha_i/2 < k < 0$, the weights of different pulses (timescales) in the total light curve are similar; these cases are classified as a group with no dominant timescale. Consequently it will show a PL distribution with a power-law index, i.e., the slope of the total PDS is $\alpha_i + 2k$ (as shown in the bottom panel of Figure 13).

The correlation between $C_i$ and $\tau_i$ may be related to energy E. The typical relation of the energy-dependent pulse width is $\tau \sim E^{-0.4}$ (Fenimore et al. 1995; Norris et al. 2005; Liang et al. 2006). $C_i$ would exhibit a power-law relation with energy within a narrow energy range (Qin 2004). The $C_i$–E relation depends on the power-law photon index $\Gamma$ (a power-law photon spectrum $N(E) \sim E^{-\Gamma}$ is assumed), then $C_i \sim E^{-(\Gamma-1)}$. We can get $k = -2.5(1-\Gamma)$ from Equation (17). For Insight-HXMT LGRBs, the $\alpha$ value of individual PDSs for the group with a dominant timescale (BPL group) is $\alpha_i = 2.49 \pm 0.79$, and the slope of the individual PDSs for the group with no dominant timescale (PL group) is $\alpha_i + 2k = 1.95 \pm 0.74$. For Insight-HXMT SGRBs, the $\alpha$ value of individual PDSs for the BPL group is $\alpha_i = 2.84 \pm 1.12$, and for the PL group it is $\alpha_i + 2k = 1.94 \pm 0.34$. Then, we roughly estimate that the value of the $\Gamma$ index for LGRBs in the PL group is $1.07 \pm 0.32$, which is consistent with the typical value of the low-energy index with a Band function; the value of the $\Gamma$ index for SGRBs in the PL group is estimated as $0.82 \pm 0.33$, which indicates that energy spectra of SGRBs are slightly harder than those of LGRBs (Preece et al. 2000; Nava et al. 2011; Zhang 2011; Goldstein et al. 2012; Bošnjak et al. 2014; Gruber et al. 2014; Song et al. 2022).

Observations have shown that the light curves of GRBs are superposed with two components. One is a fast component (e.g., < 1 s); the other is a slow component (e.g., >1 s) (Scargle 1998; Shen & Song 2003; Vetere et al. 2006; Margutti et al. 2009; Gao & Zhang 2015). The results for the dominant timescale $\tau$ show that the values for LGRBs vary from 0.18 to 12.95 s, with a typical value of 1.58 s, while the values for SGRBs vary from 0.01 to 0.08 s, with a typical value of 0.02 s. The typical $\tau$ values of both LGRBs and SGRBs are consistent with the results from the analysis of the superposed variability components in the GRB prompt emission light curves of Insight-HXMT (the second paper in the series of a comprehensive analysis of Insight-HXMT GRB data).

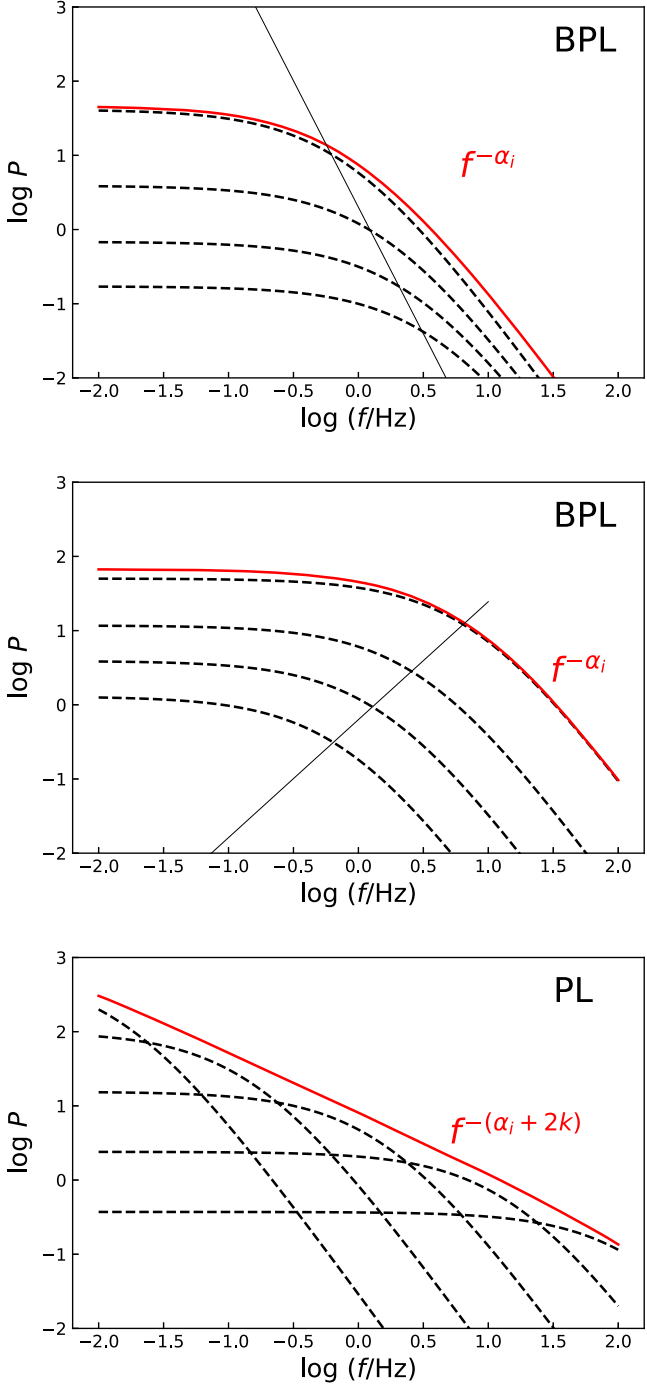

**Figure 13.** Sketch of a total PDS (red solid line) as the result of the superposition of PDSs from different pulses (black dashed lines). The black solid lines trace the break frequency $P_{b,i}(f_{b,i})$. Top panel: $C_i/C_{max} \sim \tau^k$ where $k > 0$; the overall variability is dominated by pulses with a low-frequency dominant timescale. Middle panel: $C_i/C_{max} \sim \tau^k$ where $k < -\alpha_i/2$; the overall variability is dominated by pulses with a high-frequency dominant timescale. Bottom panel: $C_i/C_{max} \sim \tau^k$ where $-\alpha_i/2 < k < 0$, and no break stands out in the total PDS, which looks like a power law with the index of $\alpha_i + 2k$.

## 5. Conclusion

Taking advantage of its large effective area, Insight-HXMT/HE provides powerful data to reveal the nature of GRB prompt emission. We perform statistical studies of GRB prompt emission with Insight-HXMT GRB data, which will be presented in a series of papers. In this paper, we focus on the PDSs. Our results can be summarized as follows.

1. For the average PDS of LGRBs, $\alpha_{\bar{P}}$ decreases as energy increases, with $1.58^{+0.06}_{-0.06}$, $1.37^{+0.07}_{-0.06}$, and $1.29^{+0.17}_{-0.17}$ for the three energy bands (100–245, 245–600, and 600–2000 keV). There is no obvious break observed in the PDS for frequencies up to ∼2500 Hz.

2. For the average PDS of SGRBs, $\alpha_{\bar{P}}$ also decreases as energy increases, with $1.52^{+0.55}_{-0.36}$ and $1.48^{+0.58}_{-0.44}$ for 100–245 and 245–600 keV, respectively. We do not observe any obvious break of average power density up to a frequency of ∼100 Hz. We also obtain the values of $\alpha_{\bar{P}}$ for Swift/BAT: $1.86^{+0.16}_{-0.13}$ and $1.71^{+0.11}_{-0.09}$ in the energy bands 15–50 and 50–150 keV, respectively. For Fermi/GBM, the values of $\alpha_{\bar{P}}$ are $1.60^{+0.18}_{-0.15}$, $1.56^{+0.07}_{-0.07}$, and $1.34^{+0.12}_{-0.11}$ in the energy bands 8–40, 40–200, and 200–1000 keV, respectively. The values of $\alpha_{\bar{P}}$ for both LGRBs and SGRBs are similar and decrease as energy increases. The similarity between the PDSs of LGRBs and SGRBs indicates that they may be the result of similar stochastic processes.

3. For the individual PDSs of LGRBs, the typical values of $\alpha$ for BPL and PL models are $2.49 \pm 0.78$ and $1.95 \pm 0.74$, respectively. The $\alpha$ value of BPL is higher than the PL one, which is similar to the results from Guidorzi et al. (2016).

4. For the individual PDSs of SGRBs, the typical values of $\alpha$ of the BPL and PL samples are $2.84 \pm 1.12$ and $1.94 \pm 0.34$, respectively, which are similar to the $\alpha$ for those of LGRBs.

5. Combining the results from Insight-HXMT and Swift/BAT, CGRO/BATSE, Fermi/GBM, and BeppoSAX/GRBM, we obtain the $\alpha_{\bar{P}} - E$ relation from 8 to 2000 keV, with $\alpha_{\bar{P}} \propto E^{-0.09 \pm 0.01}$. The Insight-HXMT data allow us to extend the energy of the $\alpha_{\bar{P}} - E$ relation to 2000 keV.

6. We obtain the relation $\alpha_{\bar{P}} - E$ of SGRBs from 8 to 1000 keV, with $\alpha_{\bar{P}} \propto E^{-0.07 \pm 0.02}$. The trend of the $\alpha_{\bar{P}} - E$ relation for SGRBs is similar to that for LGRBs.

7. For LGRBs, the typical value of the dominant timescale $\tau$ is 1.58 s, and it varies from 0.18 to 12.95 s. For SGRBs, $\tau$ ranges between 0.001 and 0.08 s, with a typical value of 0.02 s. The dominant timescale increases with increasing duration of GRBs, with $\tau \propto T_{90}^{0.86 \pm 0.09}$ for both LGRBs and SGRBs, respectively. It seems that the dominant timescale is proportional to the duration of GRBs.

8. Based on the average PDS and the individual PDSs of Insight-HXMT GRBs, no periodic or quasi-periodic signal above $3\sigma$ significance threshold is found. There are obvious different behaviors between the group with a dominant timescale (BPL model) and the group with no dominant timescale (PL model). This may be because the GRB light curves are the results of superposing a number of pulses with different timescales, which may be related to the fast and slow components.

## Acknowledgments

We thank the Insight-HXMT mission for providing valuable data, and thank the referee for very detailed comments that helped us to improve the paper significantly. This work is supported by the National Natural Science Foundation of China (grant Nos. 12373042, U1938201, 12103055, 12133003, 12133007), the Bagui Scholars Programme (W.X.-G.), the Guangxi Science Foundation (grant No. 2018GXNSFGA281007), and the Innovation Project of Guangxi Graduate Education (grant No. YSCW2019050).

## Appendix

For the light curve of a GRB, we obtained the net counts with duration $T$ (e.g., $3 \times T_{5\sigma}$). We set the binning time as $\Delta t$, and then we can obtain the net counts $C_l$ ($l = 0, ..., n-1$) in the $l$th time bin, where $n = \text{int}(T/\Delta t)$.[10] The $j$th Fourier coefficient (e.g., Leahy et al. 1983) is

$$A_j = \sum_{l=0}^{n-1} (C_l - \overline{C}) e^{2\pi i j l / n}, \quad (A1)$$

where the corresponding $j$th frequency $f_j = j/T$ ($j = 1, ..., \text{int}(n/2)$) (Cooley & Tukey 1965; Bluestein 1970), and $\overline{C}$ is the mean value of the net counts. The $j$th power density is given by (Leahy et al. 1983)

$$P_j = 2 \sum_{l,l'=0}^{n-1} (C_l - \overline{C})(C_{l'} - \overline{C}) e^{2\pi i j (l-l')/n}$$
$$= 2 \left[ \sum_{l,l'=0}^{n-1} C_l C_{l'} e^{2\pi i j (l-l')/n} - \overline{C} \sum_{l,l'=0}^{n-1} C_l (e^{2\pi i j (l-l')/n} + e^{-2\pi i j (l-l')/n}) + \overline{C}^2 \sum_{l,l'=0}^{n-1} e^{2\pi i j (l-l')/n} \right], \quad (A2)$$

where $l'$ ($l' = 0, ..., n-1$) is also the subscript of the net counts. Since

$$\sum_{l,l'=0}^{n-1} e^{2\pi i j (l-l')/n} = 0, \quad (A3)$$

and

$$\sum_{l,l'=0}^{n-1} C_l (e^{2\pi i j (l-l')/n} + e^{-2\pi i j (l-l')/n})$$
$$= 2 \sum_{l=0}^{n-1} C_l \sum_{l'=0}^{n-1} \cos[2\pi j (l-l')/n] = 0, \quad (A4)$$

Equation (A2) could be written as

$$P_j = 2 \sum_{l,l'=0}^{n-1} C_l C_{l'} e^{2\pi i j (l-l')/n}. \quad (A5)$$

---

[10] $\text{int}(x)$ is used to represent the integer part of arbitrary variable $x$ in this work.

The expected value of $P_j$ is given by

$$E\{P_j\} = 2E\left(\sum_{l,l'=0}^{n-1} C_l C_{l'} e^{2\pi i j (l-l')/n}\right)$$
$$= 2 \sum_{l,l'=0}^{n-1} E(C_l C_{l'}) e^{2\pi i j (l-l')/n}$$
$$= 2 \left[ \sum_{l,l'=0}^{n-1} E(C_l) E(C_{l'}) e^{2\pi i j (l-l')/n} \right.$$
$$\left. + \sum_{l,l'=0}^{n-1} \text{Cov}(C_l, C_{l'}) e^{2\pi i j (l-l')/n} \right], \quad (A6)$$

where $\text{Cov}(C_l, C_{l'})$ is the covariance of net counts, and $E(C_l C_{l'}) = E(C_l) E(C_{l'}) + \text{Cov}(C_l, C_{l'})$ (Rice 2007) is used.

$C_l$ is calculated by subtracting the background counts $C_{B,l}$ from the total counts $C_{O,l}$, in the $l$th bin. If $C_{B,l}$ is considered as a constant, the covariance of the original light curve, $\text{Cov}(C_{O,l}, C_{O,l'})$, is (Rice 2007)

$$\text{Cov}(C_{O,l}, C_{O,l'}) = \text{Cov}(C_l + C_{B,l}, C_{l'} + C_{B,l'})$$
$$= \text{Cov}(C_l, C_{l'}). \quad (A7)$$

On the other hand, the covariance satisfies (Rice 2007)

$$\text{Cov}(C_{O,l}, C_{O,l'}) = \begin{cases} 0, & l \neq l' \\ \sigma_l^2, & l = l', \end{cases} \quad (A8)$$

where $\sigma_l^2$ is the variance of the $l$th total counts $C_{O,l}$. For a Poisson process, $\sigma_l^2$ is equal to $C_{O,l}$, so the total variance

$$V = \sum_{l=0}^{n-1} \sigma_l^2 = \sum_{l=0}^{n-1} C_{O,l}. \quad (A9)$$

Hence, $E\{P_j\}$ could be written as

$$E\{P_j\} = 2V. \quad (A10)$$

For peak normalization, the expected value of Poisson level $B_p$ could be obtained from $E\{P_j\}$ divided by $C_p^2$ (e.g., Beloborodov et al. 1998),

$$B_p = \frac{E\{P_j\}}{C_p^2} = \frac{2V}{C_p^2}. \quad (A11)$$

The power density of a GRB light curve would tend to $\sim B_p$ at high frequency.

## ORCID iDs

Zi-Min Zhou ● https://orcid.org/0000-0003-3360-2211
Xiang-Gao Wang ● https://orcid.org/0000-0001-8411-8011
En-Wei Liang ● https://orcid.org/0000-0002-7044-733X
Cheng-Kui Li ● https://orcid.org/0000-0001-5798-4491
Bing Li ● https://orcid.org/0000-0002-0238-834X
Da-Bin Lin ● https://orcid.org/0000-0003-1474-293X
Tian-Ci Zheng ● https://orcid.org/0000-0001-6076-9522
Rui-Jing Lu ● https://orcid.org/0000-0003-2467-3608
Shao-Lin Xiong ● https://orcid.org/0000-0002-4771-7653
Ling-Jun Wang ● https://orcid.org/0000-0002-8352-1359
Li-Ming Song ● https://orcid.org/0000-0003-0274-3396
Shuang-Nan Zhang ● https://orcid.org/0000-0001-5586-1017